\documentclass{jfm}
\usepackage{graphicx}
\usepackage{epstopdf, epsfig}
\usepackage{kotex}
\usepackage{cleveref}
\usepackage{multirow}
\usepackage{booktabs}

\usepackage[dvipsnames]{xcolor}

\shorttitle{Unsupervised deep learning for super-resolution reconstruction}
\shortauthor{H. Kim, J. Kim, S. Won and C. Lee}

\title{Unsupervised deep learning for super-resolution reconstruction of turbulence}

\author{Hyojin Kim\aff{1}, Junhyuk Kim\aff{1}\footnote{Co-first author}, Sungjin Won\aff{2} \and Changhoon Lee\aff{1,2}
	\corresp{\email{clee@yonsei.ac.kr}}}
\affiliation{\aff{1}Department of Mechanical Engineering, Yonsei University, Seoul 03722, Korea
	            \aff{2}Department of Computational Science and Engineering, Yonsei University, Seoul 03722, Korea}

\begin{document}
\maketitle

\begin{abstract}
Recent attempts to use deep learning for super-resolution reconstruction of turbulent flows have used supervised learning, which requires paired data for training. This limitation hinders more practical applications of super-resolution reconstruction. Therefore, we present an unsupervised learning model that adopts a cycle-consistent generative adversarial network that can be trained with unpaired turbulence data for super-resolution reconstruction. Our model is validated using three examples: (i) recovering the original flow field from filtered data using direct numerical simulation (DNS) of homogeneous isotropic turbulence; (ii) reconstructing full-resolution fields using partially measured data from the DNS of turbulent channel flows; and (iii) generating a DNS-resolution flow field from large eddy simulation (LES) data for turbulent channel flows. In examples (i) and (ii), for which paired data are available for supervised learning, our unsupervised model demonstrates qualitatively and quantitatively similar performance as that of the best supervised-learning model. More importantly, in example (iii), where supervised learning is impossible, our model successfully reconstructs the high-resolution flow field of statistical DNS quality from the LES data. This demonstrates that unsupervised learning of turbulence data is indeed possible, opening a new door for the wide application of super-resolution reconstruction of turbulent fields. 
\end{abstract}

\begin{keywords}
	Authors should not enter keywords on the manuscript, as these must be chosen by the author during the online submission process and will then be added during the typesetting process (see http://journals.cambridge.org/data/\linebreak[3]relatedlink/jfm-\linebreak[3]keywords.pdf for the full list)
\end{keywords}

\section{Introduction}
Turbulence is a chaotic, spatio-temporal multi-scale nonlinear phenomenon. Thus, it generally requires huge costs to accurately measure or simulate with sufficiently high resolution. In particular, direct numerical simulation has been actively used in the study of turbulence. However, securing the computational resources needed to resolve even the smallest-scale motions of turbulence is progressively challenging with high Reynolds numbers. To help resolve this problem, a neural network (NN) having the capability to approximate arbitrary nonlinear functions \citep{Hornik1989} has been examined. Indeed, there have been attempts to apply NNs to the representation of turbulence \citep{Lee1997, Milano2002}. However, those applications were based on shallow learning and, thus, were restricted to the extraction of simple correlations between turbulence quantities at two close locations in a near-wall flow. In recent years, deep NNs (DNN) have been extended to various fields of turbulence research, owing to the development of data-driven learning algorithms (e.g., deep learning \citep{LeCun2015}), computational equipment (e.g., graphical process units), big data (e.g., Johns--Hopkins Turbulence Database (JHTDB) \citep{Perlman2007JHTDB}), and open-source code (e.g., TensorFlow \citep{Abadi2016}).

Various deep-learning applications have recently been developed for wide areas of turbulence research. \citet{Ling2016} proposed a tensor-based NN by embedding the Galilean invariance of a Reynolds-averaged Navier--Stokes (RANS) model, showing a greater performance improvement than linear and nonlinear eddy viscosity models. \citet{Parish2016}, \citet{Wang2017}, and other researchers have actively engaged in improving RANS models (e.g., \citet{Kutz2017,Duraisamy2019}). On the other hand, \citet{Gamahara2017} proposed a large eddy simulation (LES)-closure model based on DNN for wall-bounded turbulence. It was then extended to other flows, such as 2D turbulence \citep{Maulik2019} and homogeneous isotropic turbulence \citep{Beck2019,Xie2019}. Additionally, the prediction of the temporal evolution of turbulent flows has been actively pursued. As a fundamental example, \citet{Lee2019} studied the historical prediction of flow around a cylinder using generative adversarial networks (GAN). \citet{Srinivasan2019} predicted the temporal behavior of simplified shear turbulence expressed as solutions of nine ordinary differential equations using a recurrent NN (RNN). \citet{Kim2020} proposed a high-resolution inflow turbulence generator at various Reynolds numbers, combining a GAN and an RNN. As another noticeable attempt to apply machine learning to fluid dynamics, \citet{Raissi2020} reconstructed velocity and pressure fields from only visualizable concentration data based on a physics-informed NN framework. Recently, deep-reinforcement learning has been applied to fluid dynamics, such as observations of how swimmers efficiently use energy \citep{Verma2018} and the development of a new flow-control scheme \citep{Rabault2019}.

Apart from the above studies, the super-resolution reconstruction of turbulent flows has recently emerged as an interesting topic. This capability would help researchers overcome environments in which only partial or low-resolution spatio-temporal data are available, owing to the limitations of measurement equipment or computational resources. Particularly, if DNS-quality data could be reconstructed from data obtained via LES, it would be very helpful for sub-grid scale modeling. \citet{Maulik2017} proposed a shallow NN model that could recover a turbulent flow field from a filtered or noise-added one. \citet{Fukami2019} reconstructed a flow field from a low-resolution filtered one using a convolutional NN (CNN) for 2D decaying isotropic turbulence. \citet{Liu2020} applied temporal effects to a CNN model, showing better performance than a static model. Both \citet{Fukami2019} and \citet{Liu2020} trained CNNs in the direction of reducing the mean-squared error (MSE) of target quantities between prediction and true data. However, small-scale structures were not represented well when the resolution ratio between target and input fields was large. \citet{Deng2019} considered flow data around a cylinder measured using particle image velocimetry (PIV) in a learning network using a GAN in which the small-scale structures were better expressed than when only MSE was used. In all these prior studies, researchers used a supervised deep-learning model, which required labeled low- and high-resolution data for training. Therefore, paired data were artificially generated by filtering or averaging so that supervised learning could be made possible. In a more practical environment, however, only unpaired data are available (e.g., LES data in the absence of corresponding DNS data or measured data using PIV with limited resolution). For more practical and wider applications, a more generalized model that can be applied, even when paired data are not available, is needed. \citet{Kim2020} recently showed that unsupervised learning networks could generate turbulent flow fields for inflow boundary conditions from random initial seeds. This indeed demonstrates that a DNN can learn and reflect hidden similarities in unpaired turbulence. Based on this evidence, we presume that super-resolution reconstruction of unpaired turbulence is now possible by learning the similarities among the unpaired data.

\begin{figure}
	\centerline{\includegraphics[width=1\columnwidth]{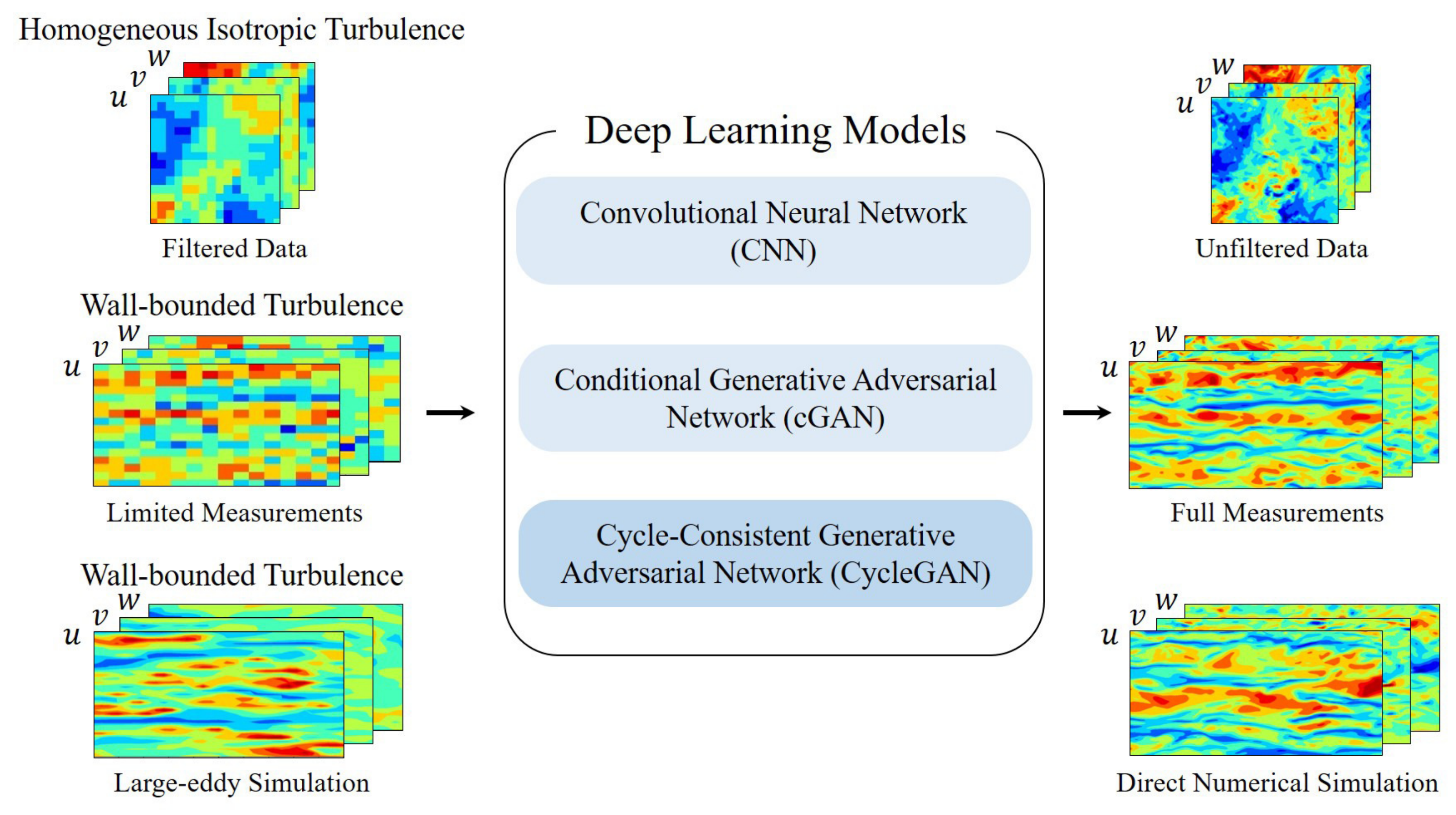}}
	\caption{Illustration of present work}
	\label{DL}
\end{figure}

In this paper, we propose an unsupervised deep-learning model that can be used, even in the absence of labeled turbulent data. For a super-resolution reconstruction using unpaired data, we apply a cycle-consistent GAN (CycleGAN) \citep{Zhu2017} to various turbulent flows as an unsupervised learning model. The detailed methodology is presented in Section \ref{sec:methodology}. For comparison, we use bicubic interpolation and supervised learning models (i.e., CNN and conditional GAN (cGAN)). The models are applied to three examples, as shown in figure \ref{DL}. First, with homogeneous isotropic turbulence, a reconstruction of the DNS flow field from a top-hat-filtered (i.e., low-resolution) one is considered in Section \ref{subsec:isotropic}. Next, in Section \ref{subsec:channel}, we cover the reconstruction of full DNS data from a partially measured (i.e., low-resolution) one in wall-bounded turbulence. In Sections \ref{subsec:isotropic} and \ref{subsec:channel}, we train our CycleGAN model using unpaired datasets with supervised learning models using paired ones. Finally, in Section \ref{subsec:LES}, DNS-quality reconstruction from LES is addressed using independently obtained LES and DNS data of wall-bounded turbulence. In this case, only the unsupervised learning model is applicable. We conclude our study with a discussion in Section \ref{Conclusion}.

\section{Methodology}\label{sec:methodology}
\begin{figure}
	\centerline{\includegraphics[width=0.8\columnwidth]{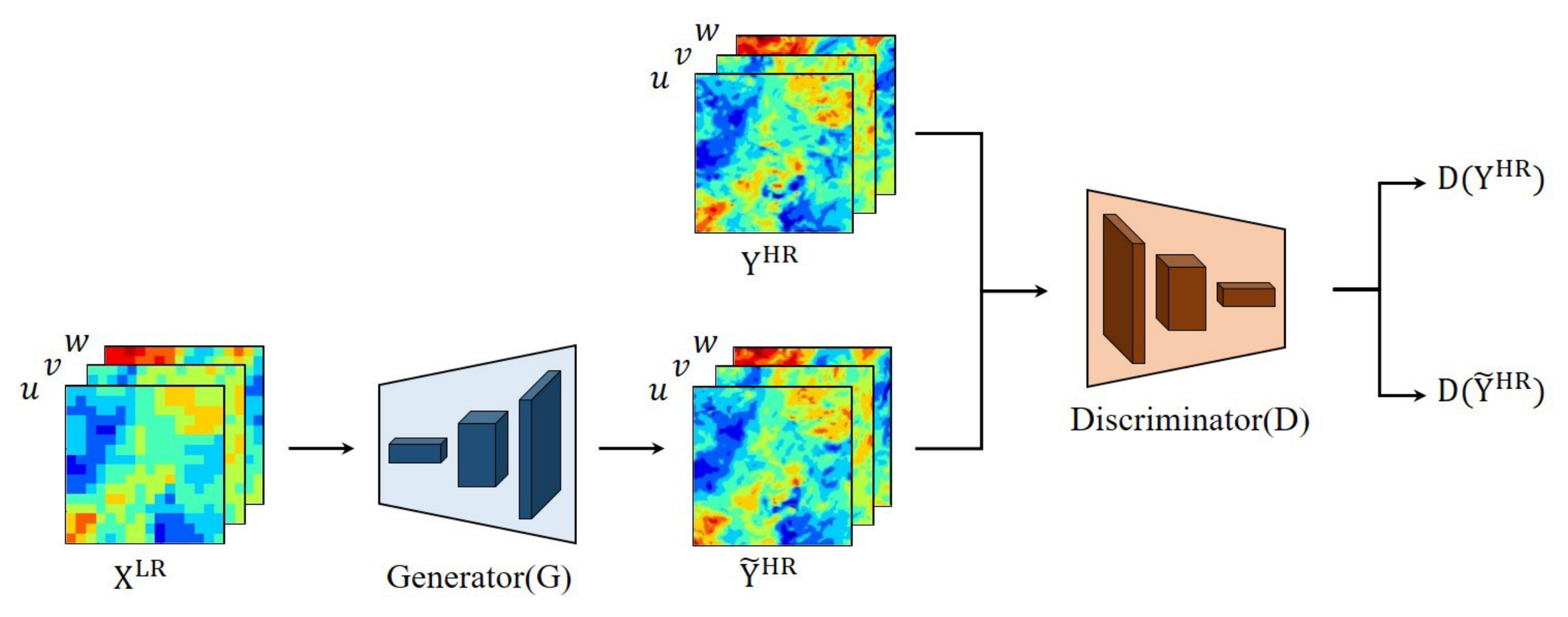}}
	\caption{GAN architecture}
	\label{GAN}
\end{figure}

In this study, we apply CycleGAN to an unsupervised learning task. A typical GAN model consists of two networks: a generator network, $(G)$, and a discriminator network, $(D)$\citep{Goodfellow2014}. In the field of image generation, $G$ generates a fake image similar to the real one by applying convolution and up-sampling to random noise $z$. $D$ distinguishes between the fake image and the real one and returns a probability value between 0 and 1 by applying convolution and down-sampling. The final goal is to obtain $G$, which can generate fake images that are difficult to distinguish from real ones. This process is similar to a min--max two-player game for the value function, $V(D,G)$, as follows:
\begin{equation}
\min_{G} \max_{D} V(D,G) = \mathbb{E}_{x\sim P_{X}}[\log D(x)] + \mathbb{E}_{z\sim P_{Z}}[\log(1-D(G(z)))],
\label{eq:GAN.loss}
\end{equation}
where $X$ is a real image set, and $x\sim P_{X}$ means that $x$ is sampled from the real image distribution. $z$ is a random noise vector of latent space used as the input to the generator. $G$ is expected to generate a fake image similar to the real one. Thus, trainable parameters in $G$ are trained in the direction of $D(G(z))$, having a value close to 1. On the other hand, those in $D$ are trained in the direction of $D(x)$, returning a value close to 1. $D(G(z))$ returns a value close to 0. Thus, even a slight difference between the real image and the generated one can be distinguished. In other words, the $G$ parameters are adjusted in a direction that minimizes $V(D,G)$, and $D$ parameters are adjusted in a direction that maximizes $V(D,G)$. From this competitive learning, we can expect to obtain a generator, $G$, capable of providing a new image having a distribution similar to a real one. In the present work, GAN is applied to super-resolution reconstruction in the frame of finding an input--output mapping function, and, instead of random noise, low-resolution image data are used as the input of $G$, as illustrated in Figure \ref{GAN}.

\begin{figure}
	\centerline{\includegraphics[width=1.0\columnwidth]{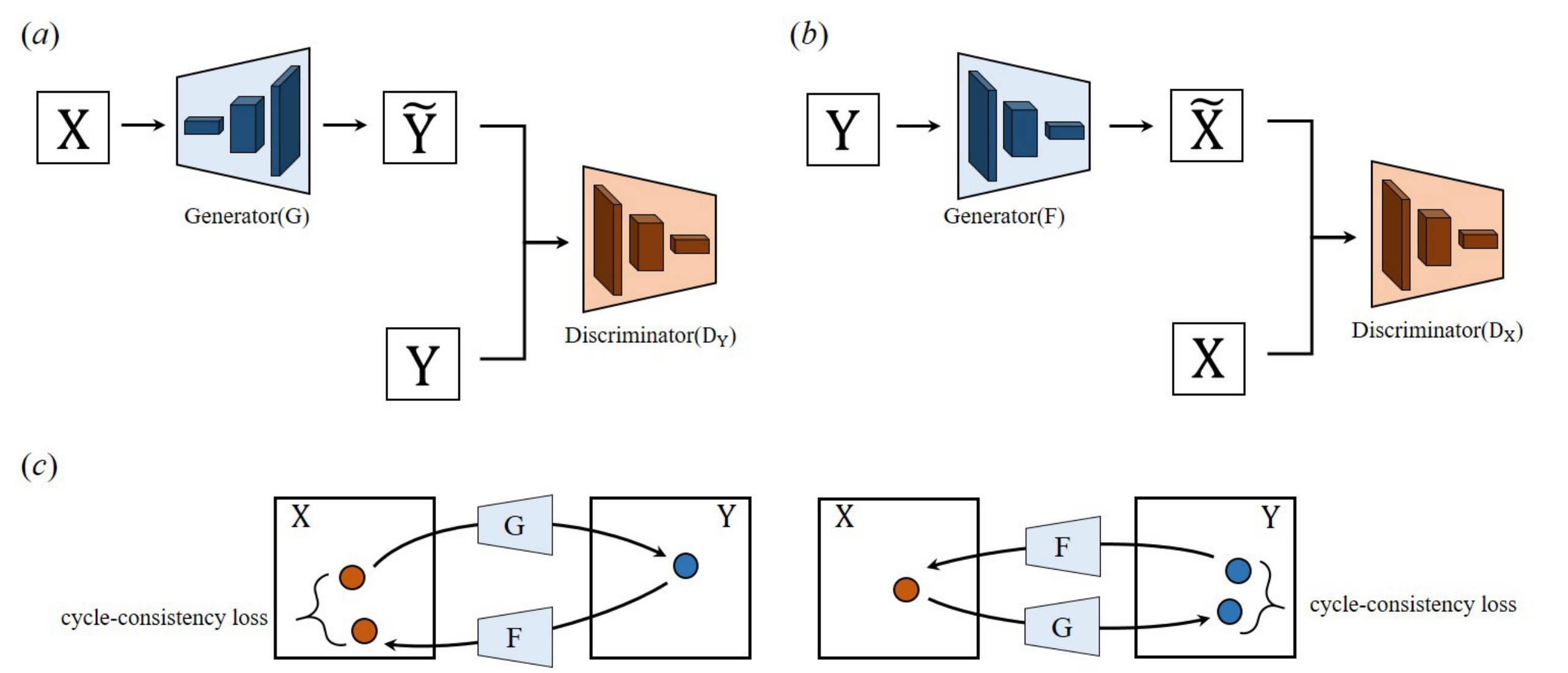}}
	\caption{ CycleGAN architecture consisting of (\textit{a}) forward GAN and (\textit{b}) backward GAN. $G$ and $F$ are generators, and $D_Y$ and $D_X$ are discriminators. (\textit{c}) Forward cycle-consistency loss: $x\rightarrow G(x)\rightarrow F(G(x)) \approx x$, and backward cycle-consistency loss: $y\rightarrow F(y)\rightarrow G(F(y)) \approx y$.}
	\label{CycleGAN}
\end{figure}

For an unsupervised learning model of unpaired turbulence, we adopt CycleGAN \citep{Zhu2017} to find a mapping function between unpaired data, $X$ and $Y$. We aim to obtain a model that performs super-resolution reconstruction when the low- and high-resolution flow fields are not matched. $X$ and $Y$ are low- and high-resolution datasets, respectively. CycleGAN consists of two generator networks, $(G, F)$, and two discriminator networks, $(D_Y, D_X)$, as shown in Figure \ref{CycleGAN}(\textit{a,b}). $G$ and $F$ are networks mapping $X \longrightarrow Y$ and $Y \longrightarrow X$, respectively. $D_Y$ and $D_X$ distinguish between a fake image from generators and a real image, returning a probability value. $D_Y$ distinguishes between $G(x)$ generated by $G$ and $y$ from $Y$, whereas $D_X$ distinguishes between $F(y)$ generated by $F$ and $x$ from $X$. The objective function of CycleGAN consists of the GAN and cycle-consistency losses. The GAN loss helps the generators find the distribution of the target image. The cycle-consistency loss connects two generators, $(G,F)$, and reflects the dependency of input on them. First, the GAN loss function is used as follows:
\begin{equation}
\mathcal{L}_{GAN}(G,D_Y) = \mathbb{E}_{y\sim P_{Y}}[\log D_Y(y)] + \mathbb{E}_{x\sim P_{X}}[\log(1-D_Y(G(x)))],
\label{eq:GAN.loss1}
\end{equation}
\begin{equation}
\mathcal{L}_{GAN}(F,D_X) = \mathbb{E}_{x\sim P_{X}}[\log D_X(x)] + \mathbb{E}_{y\sim P_{Y}}[\log(1-D_X(F(y)))],
\label{eq:GAN.loss2}
\end{equation}
where $x$ and $y$ are the images sampled from $X$ and $Y$ datasets, respectively. $G$ is trained in a direction to minimize $\mathcal{L}_{GAN}(D_Y,G)$, and discriminator $D_Y$ is trained in a direction to maximize $\mathcal{L}_{GAN}(D_Y,G)$. $F$ and $D_X$ in Equation \ref{eq:GAN.loss2} are trained in the same way.

In principle, the properly trained generators, $G$ and $F$, can provide data having a similar distributions as the target data, $Y$ and $X$. However, the above loss cannot guarantee that the generated image will be properly dependent upon the input image. In other words, the high-resolution image, $G(x)$, from the low-resolution one, $x$, could have the characteristics of target datasets, $Y$, and the reconstructed image, $G(x)$, might not have a large-scale similarity to the low-resolution one, $x$. This may reflect a dependency of input on generated data. Therefore, a cycle-consistent loss that reduces the space of the mapping function with $G$ and $F$ is additionally used (see Figure \ref{CycleGAN}(\textit{c})). This loss function consists of two terms for domains $X$ and $Y$. In the left panel of Figure \ref{CycleGAN}(\textit{c}), the forward cycle-consistency loss reduces the space of image $x$ and $F(G(x))$ in domain $X$. It makes $G(x)$ dependent upon $x$ ($x \rightarrow G(x) \rightarrow F(G(x)) \approx x$). Similarly, in the right panel of Figure \ref{CycleGAN}(\textit{c}), the backward cycle-consistency loss reduces the space of image $y$ and $G(F(y))$ in domain $Y$ and makes $F(y)$ dependent upon $y$ ($y \rightarrow F(y) \rightarrow G(F(y)) \approx y$). The cycle-consistency losses can be expressed as
\begin{equation}
\mathcal{L}_{cycle}(G,F) =\mathbb{E}_{x\sim P_{X}}[\parallel F(G(x)) - x\parallel^2_2] + \mathbb{E}_{y\sim P_{Y}}[\parallel G(F(y)) - y\parallel^2_2],
\label{eq:CycleGAN.loss}
\end{equation}
where the first term on the right-hand side is the forward cycle-consistency loss, and the second term is the backward cycle-consistency loss. $\parallel \parallel^2_2$ denotes mean-squared error, which is normalized by vector size. The MSE between $F(G(x))$ and $x$ and that between $G(F(y))$ and $y$ are used. The cycle-consistency loss provides a decisive effect on learning the unpaired data.
The final objective function used in this study is as follows:
\begin{equation}
\mathcal{L}(G,F,D_Y,D_X) = \mathcal{L}_{GAN}(G,D_Y) + \mathcal{L}_{GAN}(F,D_X) + \lambda \mathcal{L}_{cycle}(G,F),
\label{full loss}
\end{equation} 
where $\lambda$ is a weight factor and is fixed at $10$. Generators $G$ and $F$ are trained in the direction of minimizing $\mathcal{L}(G,F,D_Y,D_X)$, whereas discriminators $D_Y$ and $D_X$ are trained in the direction of maximizing $\mathcal{L}(G,F,D_Y,D_X)$. Learning with the above GAN loss often diverges, because the discriminator easily distinguishes between the generated image and the target one before parameters in the generator are sufficiently trained. Additionally, there is a well-known problem (i.e., mode collapse) in which the generation distribution is restricted to a small domain, although training does not diverge. To solve this problem, we change the above GAN loss to a Wasserstein GAN (WGAN) having a gradient penalty (GP) loss \citep{Gulrajani2017}. With the WGAN-GP loss, the GP term is added, and the probabilistic divergence between the real image and the generated one becomes continuous with respect to the parameters of the generator. Training and performance can, therefore, be stabilized and improved.

To effectively handle the spatial structures of turbulence, a CNN comprising discrete convolution operations and nonlinear functions is used as generators $G$ and $F$ and discriminators $D_Y$ and $D_X$, respectively. To change the dimension of the image (i.e., the flow field), up- and down-sampling are applied to generators $G$ and $F$, respectively. Down-sampling is used for discriminators $D_X$ and $D_Y$. Additionally, the fully-connected layer is used in the last two layers for the discriminators. As a nonlinear function, a leaky rectified linear unit ReLU is used:
\begin{equation}
f(x) = \left\{
\begin{array}{ll}
x, & x\geq0 \\[2pt]
\alpha x, & x<0
\end{array} \right.
\label{LeakyReLU}
\end{equation}
where $\alpha = 0.2$. This nonlinear function reliably updates the weight by avoiding the dead-ReLU problem that produces an output, $0$, for the negative input. Detailed hyperparameters used for training and network architecture are provided in Appendix \ref{appA}. For implementation, we use the TensorFlow open-source library \citep{Abadi2016}.

To assess our unsupervised learning, we consider supervised learning that adopts CNN and a cGAN. Their generators comprise the same network as does $G$ in the CycleGAN. The CNN is trained with the MSE that represents the pixel loss between the target flow field and the reconstructed one. With an L2 regularization added to prevent overfitting, the objective function of the CNN consists of the sum of MSE and L2 regularization loss, as follows:
\begin{equation}
\mathcal{L}_{CNN} =\mathbb{E}_{x\sim P_{X}}[\parallel G(x) - y\parallel^2_2]   +\frac{\lambda}{2} \sum_{k} w_k^2,
\label{eq:CNN.loss}
\end{equation}
where, in the MSE of the data sampled during training, $y$ and $G(x)$ are the DNS flow field and the predicted one, respectively. The second term is the L2 regularization loss, where $w$ represents trainable weights. $\lambda$ denotes the strength of the regularization, fixed at 0.0001. The CNN is trained in the direction of minimizing $\mathcal{L}_{CNN}$ to accurately predict the target flow field.

cGAN, as proposed by \citet{Mirza2014}, is similar to GAN. the cGAN model applies the generator input as a condition to the discriminator to constrain the output of the generator to be dependent upon the input. In this study, the dependency of low-resolution data is effectively reflected in the reconstruction of high-resolution data using low-resolution data as the condition. Thus, the correlation between the large-scale structure and the reconstructed small-scale structures of turbulence can be more accurately represented. The objective function of the cGAN is as follows:
\begin{equation}
\mathcal{L}_{cGAN} = \mathbb{E}_{y\sim P_{Y}}[\log D(y|x)] + \mathbb{E}_{x\sim P_{X}}[\log(1-D(G(x)|x))],
\label{eq:cGAN.loss}
\end{equation}
where $x$ and $y$ are sampled low- and high-resolution turbulent flow fields, respectively. A low-resolution field is used as the input of the discriminator in addition to the high-resolution one ($y$ or $G(x)$). For example, flow-field information, comprising a total of six channels, including high-resolution velocity vector fields and paired low-resolution fields, are used as input. Note that we can use cGAN only when paired data are provided.

In this study, the unpaired low- and high-resolution turbulent fields are used when training the CycleGAN, whereas the paired data are used when training the CNN and the cGAN. In the first two examples,(Sections, \ref{subsec:isotropic} and \ref{subsec:channel}), paired data exist, because low-resolution data are obtained from high-resolution DNS data. When learning the CycleGAN, low- and high-resolution data are shuffled and unpaired intentionally. In Section \ref{subsec:LES}, LES and DNS data are unpaired naturally. Thus, we cannot train the CNN and the cGAN, whereas we can train the CycleGAN in the same way as explained in Sections \ref{subsec:isotropic} and \ref{subsec:channel}. 

\section{Results and discussion}

\subsection{Example 1: filtered homogeneous isotropic turbulence}\label{subsec:isotropic}
In this section, using various resolution ratios, super-resolution reconstruction leveraging both supervised and unsupervised learning are considered for homogeneous isotropic turbulence at a Taylor-scale Reynolds number, $Re_ \lambda = 418$. Data were obtained from the JHTDB. The governing equations were incompressible Navier--Stokes equations. DNS was performed based on the pseudo-spectral method, and the domain and mesh size were $2\pi\times2\pi\times2\pi$ and $1024\times1024\times1024$, respectively. Details are given in \citet{Perlman2007JHTDB} and \citet{Li2008JHTDB}. We used 200 fields with $\Delta t = 0.02$ for training and 10 fields with $\Delta t = 0.2$ for validation. For testing, 10 fields with $\Delta t= 0.2$ were used independently of training data. In the current study, we restricted our scope to the reconstruction of 2D fields of 3D turbulent fields to confirm the plausibility of reconstructing turbulence using an unsupervised learning. Input and output data were 2D velocity fields ($u, v, w$) in an $x-y$ plane. Low-resolution velocity fields and filtered DNS (fDNS) data were obtained by applying down-sampling and average pooling (i.e., top-hat filter) to high-resolution DNS data. Average pooling is a local average operation that extracts the mean value over some area of the velocity fields. The size of DNS data was $ N_x \times N_y $, and that of the low resolution was $N_x / r \times N_y / r $, where $r$ is the resolution ratio. We considered three cases: $r=4$, $8$, and $16$. For training, the target (high-resolution) size was fixed at $N_x \times N_y = 128\times128$, which was a sub-region extracted from the training fields. This choice of input and target-domain sizes was made based on our observation that the domain length of $128 \Delta x (= 0.785)$ was greater than the integral length scale of the longitudinal two-point velocity autocorrelation of 0.373. This condition is an important guideline for the choice of the input domain, because high-resolution data at any point in the same domain can be reconstructed restrictively based on all of the data in the input domain.

To demonstrate the performance of unsupervised learning using CycleGAN for the super-resolution reconstruction of turbulent flows, we tested a bicubic interpolation and two kinds of supervised learning by adopting CNN and cGAN. Bicubic interpolation is a simple method of generating high-resolution images through interpolation using data at 16 adjacent pixels without learning. CycleGAN was trained using unpaired fDNS and DNS fields, and CNN and cGAN were trained using paired fDNS and DNS fields. Three velocity components, $u,v,w$, were trained simultaneously. The same hyperparameters, except those of the network architecture, were used for each resolution ratio, $r$.
\begin{figure}
	\centerline{\includegraphics[width=1.0\columnwidth]{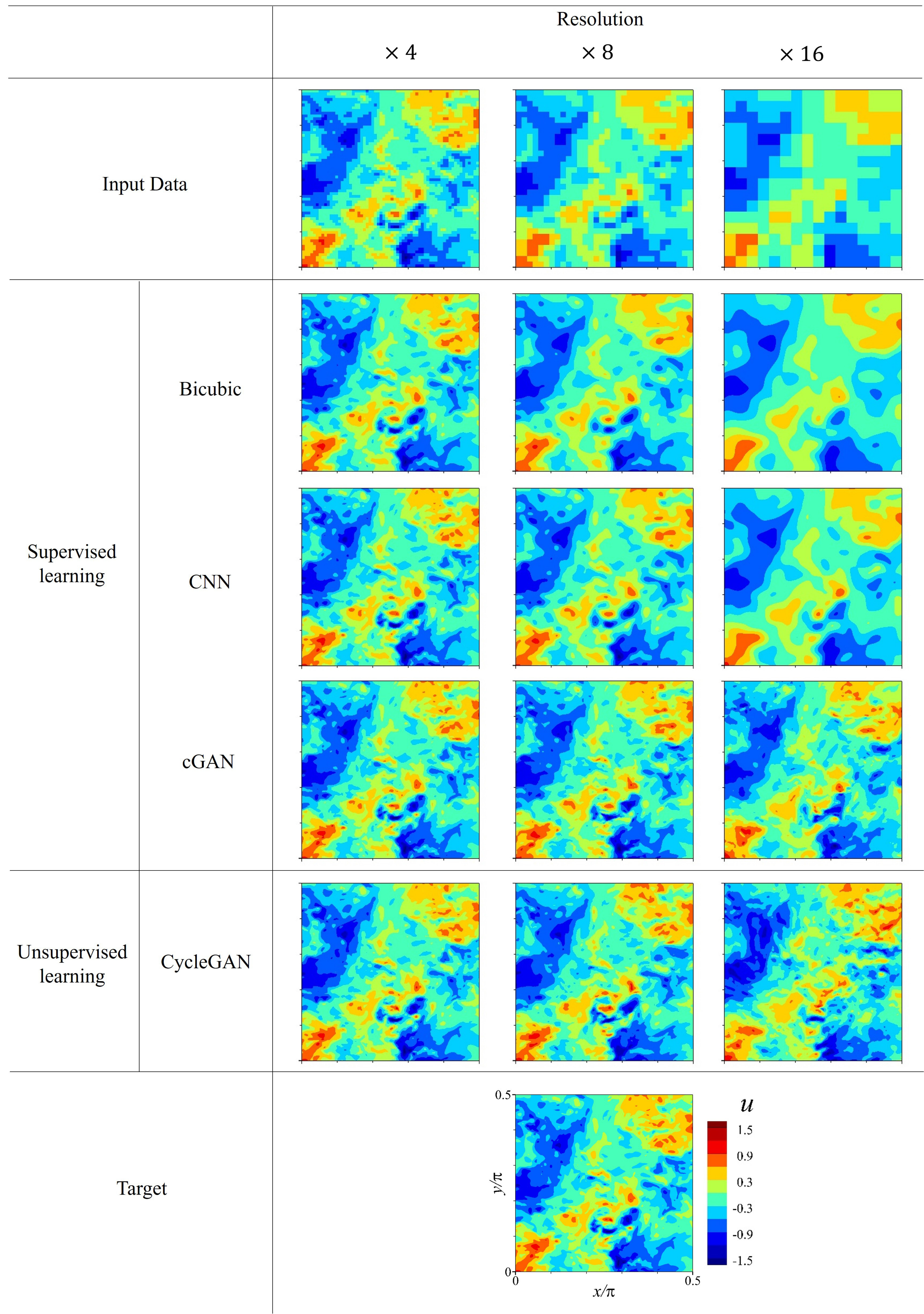}}
	\caption{Reconstructed instantaneous velocity field $(u)$ obtained by various deep-learning activities from a given low-resolution input field in the homogeneous isotropic turbulence.}
	\label{iso.velocity}
\end{figure}
The velocity, $u$, of the reconstructed 2D field, using the test data, is presented in Figure \ref{iso.velocity}. Bicubic interpolation tends to blur the target turbulence and thus cannot well-reconstruct the small scales of the target flow field, regardless of resolution ratio. This obviously indicates that the bicubic interpolation is unsuitable for small-scale reconstruction of turbulence. However, data-driven approaches can fairly well-reconstruct small-scale structures that are not included in the input data. CNN can reconstruct a velocity field similar to that of the target data when $r=4$. As $r$ increases, the CNN shows only slight improvement over bicubic interpolation. Meanwhile, cGAN can generate high-quality velocity fields similar to the DNS ones, regardless of input data resolution. 

As also shown in Figure \ref{iso.velocity}, CycleGAN showed excellent performance in reconstructing the velocity field, reflecting the characteristics of the target, given that it used unsupervised learning. When $r=4$ and $8$, our model produced a flow field quite similar to that of the target and that of the cGAN reconstruction trained using paired data. When $r=16$, the generated field by CycleGAN had a slightly different point-by-point value from the target. However, our model showed similar performance as cGAN. We used MSE to rigorously compare the difference between the target and the reconstructed flow field, as shown in Table \ref{iso.mse}. The CNN had the lowest error, whereas cGAN and CycleGAN had relatively high errors for all $r$. The reason is that the CNN model was trained in the direction of minimizing only the MSE during the training process. Other learning models that adopted GAN were trained by minimizing more sophisticated loss for proper purposes. However, as we confirm in Figure \ref {iso.velocity}, cGAN and CycleGAN had superior ability to reconstruct small scales, compared with CNN. It appears that MSE was not suitable to measure performance for super-resolution reconstruction of turbulent flows, because a bicubic interpolation produces a smaller MSE than does cGAN and CycleGAN.

\begin{table}
	\begin{center}
		\begin{tabular}{cccccc}
			\multirow{2}{*}{$r$}  & \multicolumn{4}{c}{Deep learning models}  \\ 
			\cmidrule(lr){2-5}& Bicubic& CNN & cGAN & CycleGAN \\ \hline
			4 & 0.00254 & 0.00168 & 0.00230 & 0.00548\\[1mm]
			8  & 0.01387 & 0.01019 & 0.01840 & 0.02672 \\[1mm]
			16  & 0.04140 &0.03539 & 0.06440 & 0.08677 \\[2mm]    			
		\end{tabular}
		\caption{MSE of generated velocity field for the resolution ratio, $r$. The velocity is normalized using the standard deviation of the DNS field.}{\label{iso.mse}}
	\end{center}
\end{table}

\begin{figure}
	\centerline{\includegraphics[width=1.0\columnwidth]{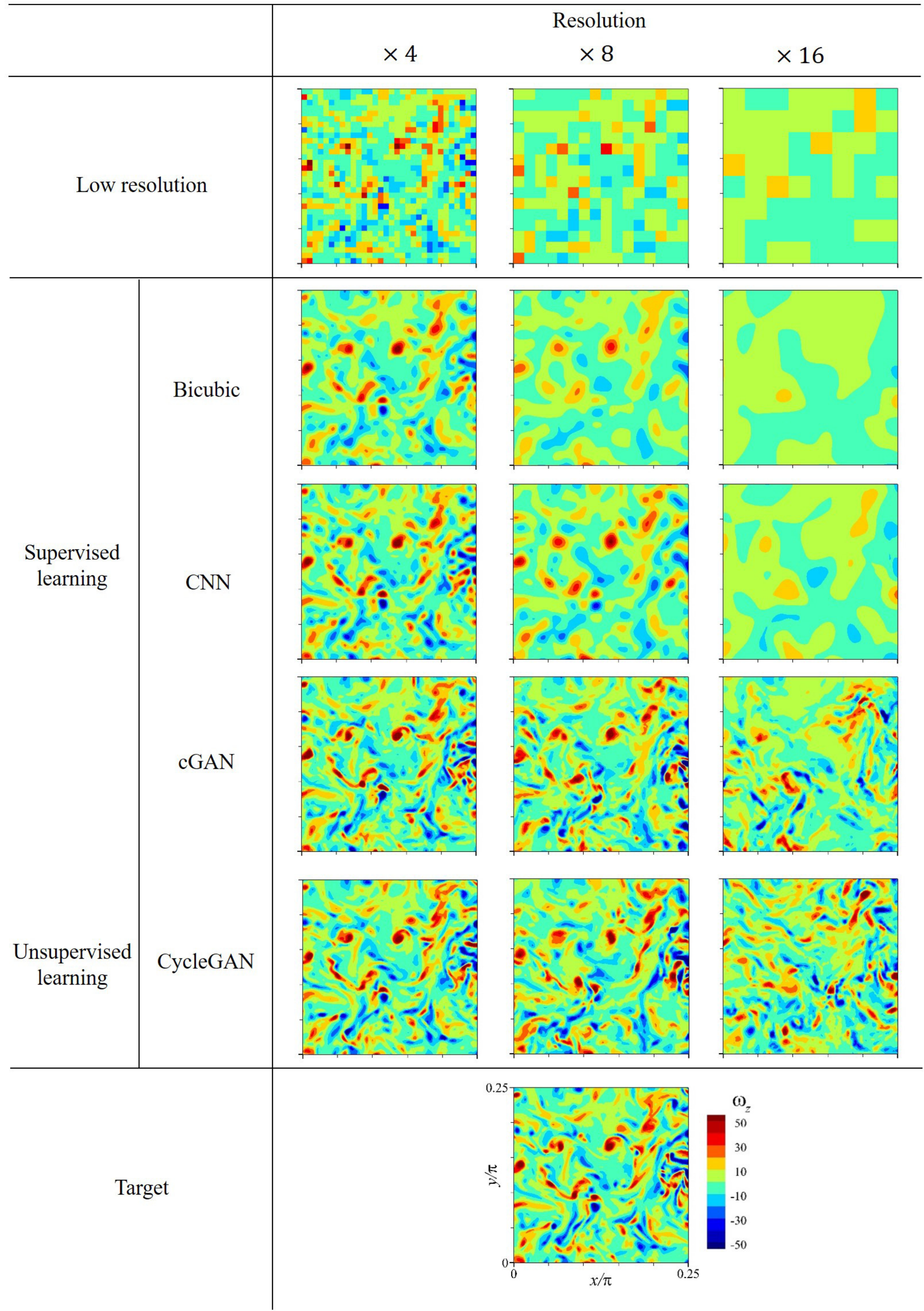}}
	\caption{Vorticity field calculated from the reconstructed velocity fields obtained by various deep-learning models.}
	\label{iso.vorticity}
\end{figure}
Vorticity field $(\omega_z)$, obtained from the reconstructed velocity information, is presented in Figure \ref{iso.vorticity}. Vorticity was not directly considered during the training process. Similar to velocity fields, bicubic interpolation and CNN were unable to reconstruct vorticity structures shown in the DNS, because the resolution of the input data decreased. However, both cGAN and CycleGAN generated vorticity structures similar to the DNS ones. However, performance was a bit deteriorated when $r =16$. 

\begin{figure}
	\centerline{\includegraphics[width=1.0\columnwidth]{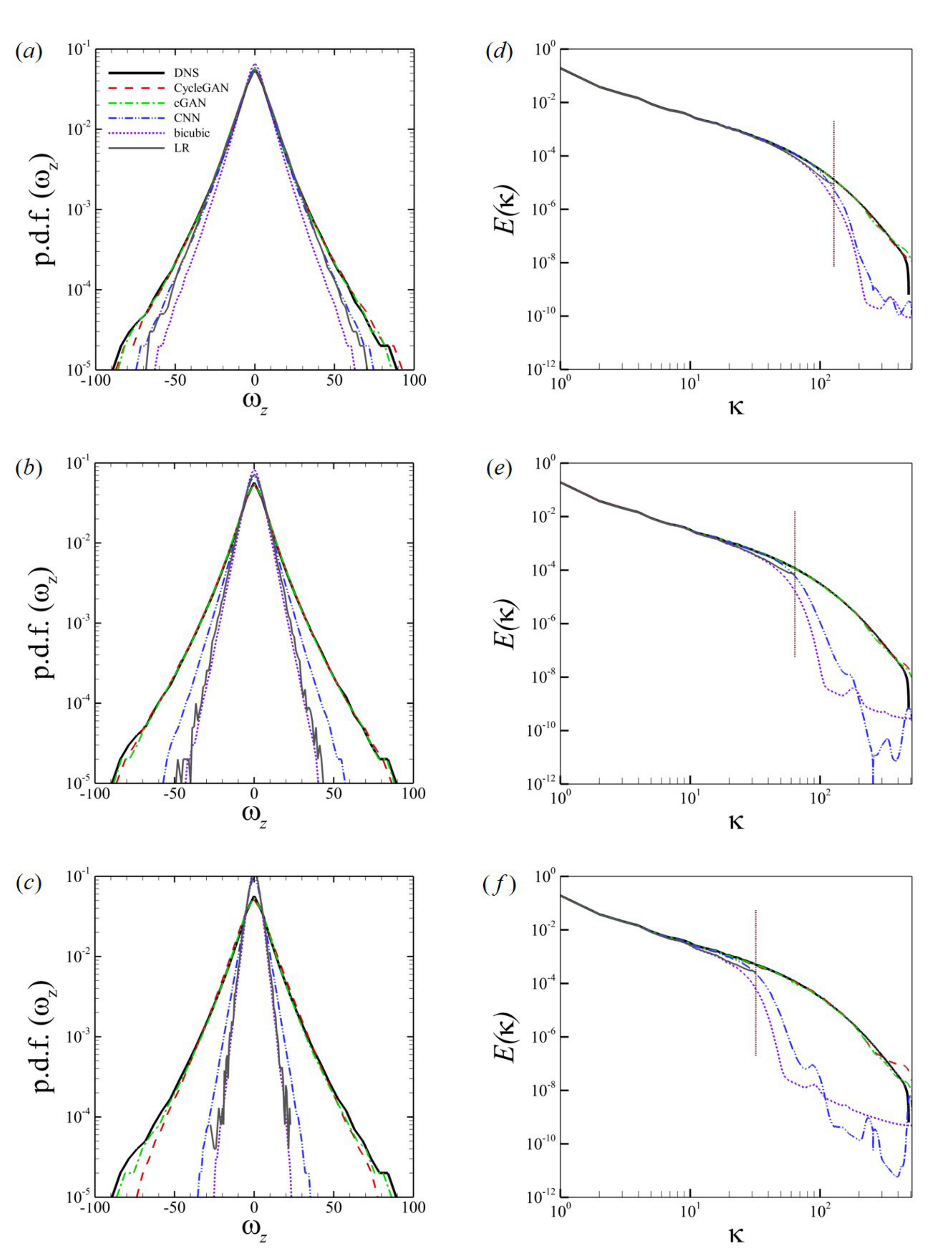}}
	\caption{Probability density function of vorticity and transverse energy spectra for various resolution ratio, $r$. (\textit{a}), (\textit{b}) and (\textit{c}) are $p.d.f.$ of vorticity corresponding to $r=4$, 8, and 16, respectively. (\textit{d}), (\textit{e}) and (\textit{f}) are energy spectra for $r=4$, 8, and 16.}
	\label{iso.energy}
\end{figure}

For more quantitative assessment of the performance of learning models, the probability density function of vorticity, $p.d.f.(\omega_z)$, for three resolution ratios are given in Figure \ref{iso.energy}(\textit{a}), (\textit{b}) and (\textit{c}). For obvious reasons, bicubic interpolation could not produce a wider distribution of the $p.d.f.$ of vorticity for DNS data, and CNN performed very poorly. On the other hand, the $p.d.f.$ of cGAN and CycleGAN recovered the DNS well, regardless of $r$. The performance of learning models in representing small-scale structures of turbulence can be better investigated using an energy spectrum. The $x$-directional energy spectrum is defined as follows:
\begin{equation}
{E(\kappa_x) = {1\over{2\pi}}\int_{-\infty}^{\infty}{e^{-ip\kappa_x}R_{V_iV_i}(p)dp}},
\label{eq.energy spectrum}
\end{equation}
where 
\begin{equation}
{{R_{V_iV_i}(p) = {\left<V_i(x,y)V_i(x+p,y) \right>}}}.
\end{equation}
Here, $\left< \right>$ denotes an average operation, and $V_i$ represents the velocity components. $R_{V_iV_i}(p)$ is the $x$-directional two-point correlation of velocity. The transverse energy spectrum is obtained by the average of the $y$-directional spectrum of $u$, the $x$-directional spectrum of $v$, and the $x$- and $y$-directional spectra of $w$. The transverse energy spectra for $r=4$, 8 and 16 are presented in Figure \ref{iso.energy} (\textit{d}), (\textit{e}) and (\textit{f}), respectively. The vertical dotted line indicates the cutoff wave number, which is the maximum wave number of low-resolution fields. Bicubic interpolation and CNN cannot represent the energy of wave numbers higher than the cutoff one. However, cGAN and unsupervised CycleGAN show great performance in recovering the energy of the DNS in the high-wave number regions, which is not included in the input data.

Test results in this section clearly indicate that CycleGAN is an effective model for super-resolution reconstruction of turbulent flows when low- and high-resolution data are unpaired. The CycleGAN model can provide statistically accurate high-resolution fields for various resolution ratios. Reconstructed velocities are very similar to targets at all $r$. Although training with unpaired data, CycleGAN performs nearly equally to cGAN, showing the best performance among supervised learning models. It appears that repetitive convolution operations and up- or down-sampling of turbulence fields in the generator and discriminator capture the essential characteristics of turbulence, which are otherwise difficult to describe.

\subsection{Example 2: measured wall-bounded turbulence}\label{subsec:channel}
To evaluate the performance of our model for anisotropic turbulence, in this section, we attempt a high-resolution reconstruction of low-resolution data for wall-bounded flows. This time, the low-resolution data were extracted from high-resolution DNS data from point-wise measurement at sparse grids instead of the local average. This is similar to experimental situations in which PIV measurements had limited spatial resolution. We used JHTDB data collected through DNS of turbulent channel flows for solving incompressible Navier--Stokes equations. The flow was driven by the mean pressure gradient in the streamwise $(x)$ direction, and a no-slip condition was imposed on the top and bottom walls. Periodicity was imposed in the streamwise, $x$, and spanwise, $z$, directions, and a non-uniform grid was used in the wall-normal direction, $y$. Detailed numerical methods were provided in \citet{Graham2015}. The friction Reynolds number, $Re_\tau=u_\tau \delta / \nu$, was defined by the friction velocity $u_\tau$, channel half-width $\delta$, and kinetic viscosity $\nu$ is 1,000. Velocity and length were normalized by $u_\tau$ and $\delta$, respectively, and superscript ($+$) was a quantity non-dimensionalized with $u_\tau$ and $\nu$. The domain length and grid resolution were $L_x \times L_y \times L_z = 8\pi \delta \times 2 \delta \times 3\pi \delta$ and $N_{x} \times N_{y} \times N_{z}$ = $2048\times 512\times1536$, respectively. The simulation time step, $\Delta t$, which was non-dimensionalized by $u_\tau$ and $\delta$, was $6.5 \times 10^{-5}$. The learning target was the streamwise velocity, $u$, the wall-normal velocity, $v$, and the spanwise velocity, $w$, in the $x-z$ plane at $y^+=15$ and $y^+=100$. $y^+=15$ is the near-wall location with maximum fluctuation intensity of $u$, and $y^+=100$ ($y / \delta=0.1$) is in the outer-region. For training and validation data, 100 fields separated by an interval, $\Delta t = 3.25 \times 10^{-3}$, and 10 fields separated by $\Delta t = 3.25 \times 10^{-2}$ were used, respectively. After training, we verified the trained model using 10 fields separated by an interval of $\Delta t = 3.25 \times 10^{-2}$ as test data. This is because they were far enough from the training data. Low-resolution partially measured data were extracted at eight-grid intervals in the streamwise and spanwise directions in the DNS fully measured data. Similar to the previous learning example in Section \ref{subsec:isotropic}, during training, input and target sizes were fixed at $16\times16$ and $128\times128$, respectively. They were sub-region extracted from training fields. Here, the streamwise input domain length was $128\Delta x=1.57$, which was greater than the integral length scale of the two-point correlation of the streamwise velocity, 1.14.

In this example, because the low-resolution data were point-wise accurate, reconstruction implies the restoration of data in-between grids where low-resolution data are given. Therefore, a stabler model can be obtained by utilizing the known values of the flow field during reconstruction. To account for the known information, a new loss term (i.e., pixel loss) is added to the existing loss function (see Equation $\ref{full loss}$). The pixel-loss function used in the unsupervised learning model, CycleGAN, is expressed as
\begin{equation}
\mathcal{L}_{pixel} =  \lambda\mathbb{E}_{x\sim P_{X}}[\frac{1}{N_{p}}\sum_{i=1}^{N_p}(x^{LR}(p_i)-y^{DL}(p_i))^2],
\label{eq:pixel loss1}
\end{equation}
where $y^{DL}$ is the reconstructed velocity field, and $x^{LR}$ is the low-resolution one. $p_{i}$ is a measured position, and $N_p$ is the number of measured points. $\lambda$ is a weight value, and we fix it to $10$. CycleGAN is trained to minimize $\mathcal{L}_{pixel}$. Table \ref{channel.error} shows the error of the test dataset, depending on the use of the pixel loss. When the pixel loss is used, the smaller error occurs at the position where exact values are known. Thus, the entire error of the reconstructed field becomes small. In the situation where a partial region is measured, a simple pixel loss could improve reconstruction accuracy for entire positions in addition to measured ones. The point-by-point accuracy can be further improved through the fine tuning of $\lambda$.
\begin{table}
	\begin{center}
		\begin{tabular}{cccccc}
			\multirow{2}{*}{}  & \multicolumn{2}{c}{Without $\mathcal{L}_{pixel}$} & \multicolumn{2}{c}{With $\mathcal{L}_{pixel}$} \\ 
			\cmidrule(lr){2-3} \cmidrule(lr){4-5} & pixel error & entire error & pixel error & entire error \\ \hline
			$y^+=15$ & 1.403 & 1.390 & 0.124 & 0.595  \\[1mm]
			$y^+=100$ & 0.871 & 0.768 & 0.075 & 0.477  \\[2mm] 			
		\end{tabular}
		\caption{Error of measured positions (i.e., pixel error) and error of entirety (i.e., entire error) for CycleGAN with and without pixel loss. The error is normalized by the standard deviation of the velocity of DNS.}{\label{channel.error}}
	\end{center}
\end{table}
\begin{figure}
	\centerline{\includegraphics[width=0.85\columnwidth]{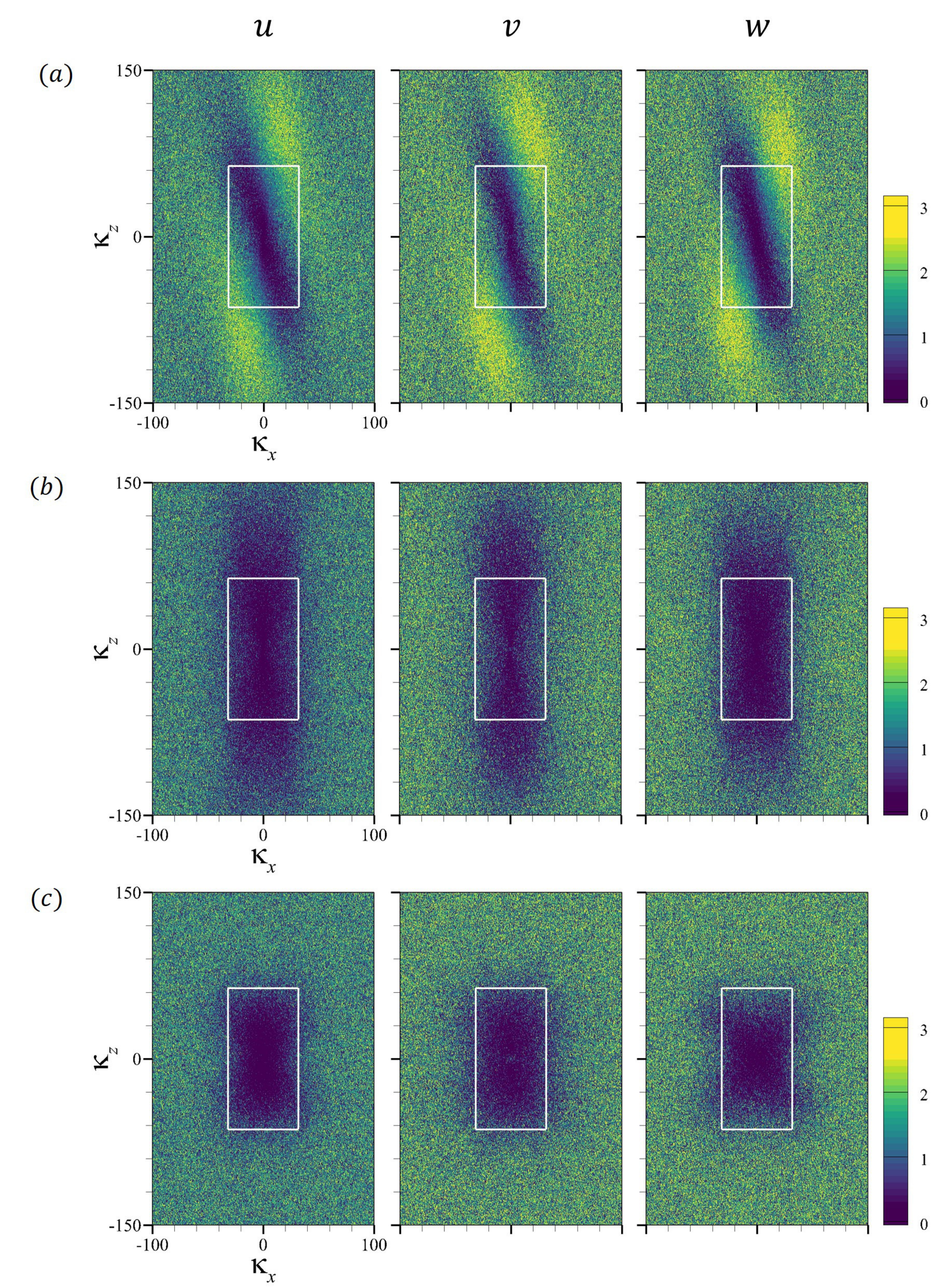}}
	\caption{Phase error defined by $|$phase$(\hat{u}_{i}^{CycleGAN})-$phase$(\hat{u}_{i}^{DNS})|$ between the Fourier coefficients of the DNS field and the generated one by CycleGAN. (\textit{a}) CycleGAN without pixel loss at $y^+=15$, (\textit{b}) CycleGAN with pixel loss at $y^+=15$, and (\textit{c}) CycleGAN with pixel loss at $y^+=100$. The box with the white line denotes the range of low-resolution input fields, $|\kappa_x|\leq \kappa_{x,cutoff}$ and $|\kappa_z| \leq \kappa_{z,cutoff}$.}
	\label{channel.phase}
\end{figure}

The absolute phase error of the Fourier coefficients in an instantaneous flow field reconstructed from test data using CycleGAN is given in Figure \ref{channel.phase}. The phase error obtained without pixel loss at $y^+=15$, that with pixel loss at $y^+=15$, and that with pixel loss at $y^+=100$ are shown in Figure \ref{channel.phase}(\textit{a}), (\textit{b}), and (\textit{c}), respectively. The maximum wave numbers of the low-resolution field, $\kappa_{x,cutoff}$ and $\kappa_{z,cutoff}$, are indicated by a white line. When pixel loss was not used, a large phase error and a phase-shift occurred for specific-size structures. This happened, because, when spatially homogeneous data are used for unsupervised learning, the discriminator cannot prevent the phase shift of high-resolution data. On the other hand, when pixel loss is used for training, the phase of all velocity components ($u, v, w$) is accurate in the area satisfying $\kappa \leq \kappa_{cutoff}$. This means that, although the large-scale structures located in the low-resolution field were well captured, the small-scale structures were reconstructed. We also noticed that the reconstructed flow field near the wall ($y^+=15$) had higher accuracy than that away from the wall ($y^+=100$) in the spanwise direction (as shown in Figure \ref{channel.phase}(\textit{b}) and (\textit{c})). This might be related to the fact that the energy of the spanwise small scale was larger in the flow field near the wall. Furthermore, as shown in Figure \ref{channel.phase}(\textit{b}), the streamwise velocity, $(u)$, at $y^+=15$ had a higher-phase accuracy in the spanwise direction compared with other velocity components, $(v, w)$. The reason might be that the energy of the streamwise velocity in high spanwise wave numbers was higher. Overall, the higher the root-mean-square (RMS) of fluctuation, the higher the phase accuracy of the reconstructed flow field.

\begin{figure}
	\centerline{\includegraphics[width=1.0\columnwidth]{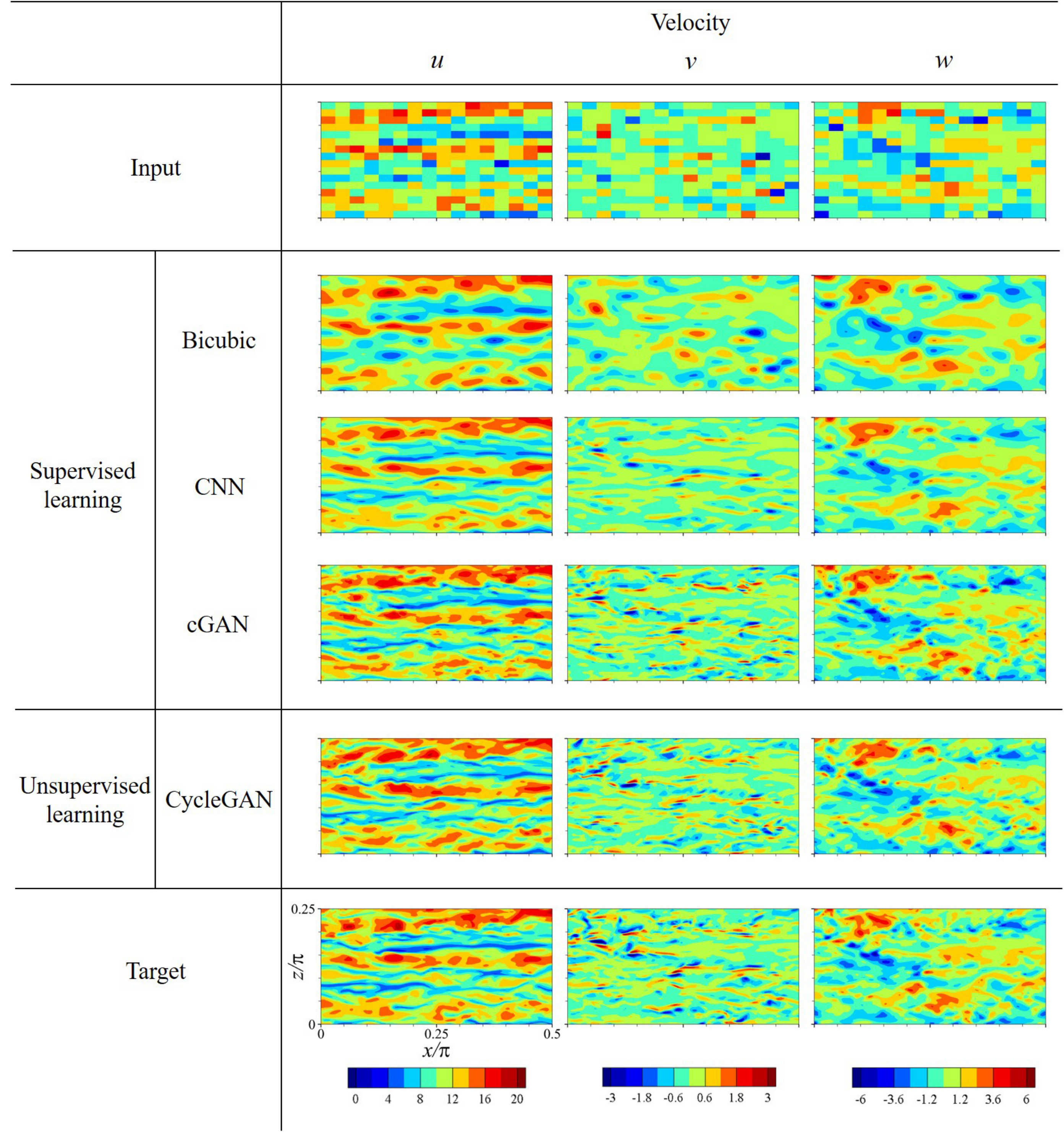}}
	\caption{Reconstructed instantaneous velocity field at $y^+=15$ obtained by various deep learning models.}
	\label{channel.velocity}
\end{figure}
Figure \ref{channel.velocity} shows the velocity field ($u, v, w$) reconstructed by various deep-learning processes from partially measured test data at $y^+=15$. For CycleGAN, an unsupervised learning model, the network was trained using unpaired data with pixel loss, and the supervised learning models (i.e., CNN and cGAN) were trained using paired data with the loss function presented in Section \ref{sec:methodology}. Bicubic interpolation smoothed the low-resolution data. Thus, it could not at all capture the characteristics of the wall-normal velocity of the DNS (target). CNN yielded slightly better results, but it had limitations in generating a flow field that reflected small-scale structures observed in the DNS field. On the other hand, cGAN demonstrated excellent capability to reconstruct the flow field, including features of each velocity value. It accurately produced a wall-normal velocity, where small scales were especially prominent. CycleGAN, an unsupervised learning model, showed similar results as cGAN, although unpaired data were used. CycleGAN reconstructed both streak structures of the streamwise velocity and small strong structures of the wall-normal velocity, similar to the DNS field.

\begin{figure}
	\centerline{\includegraphics[width=1.0\columnwidth]{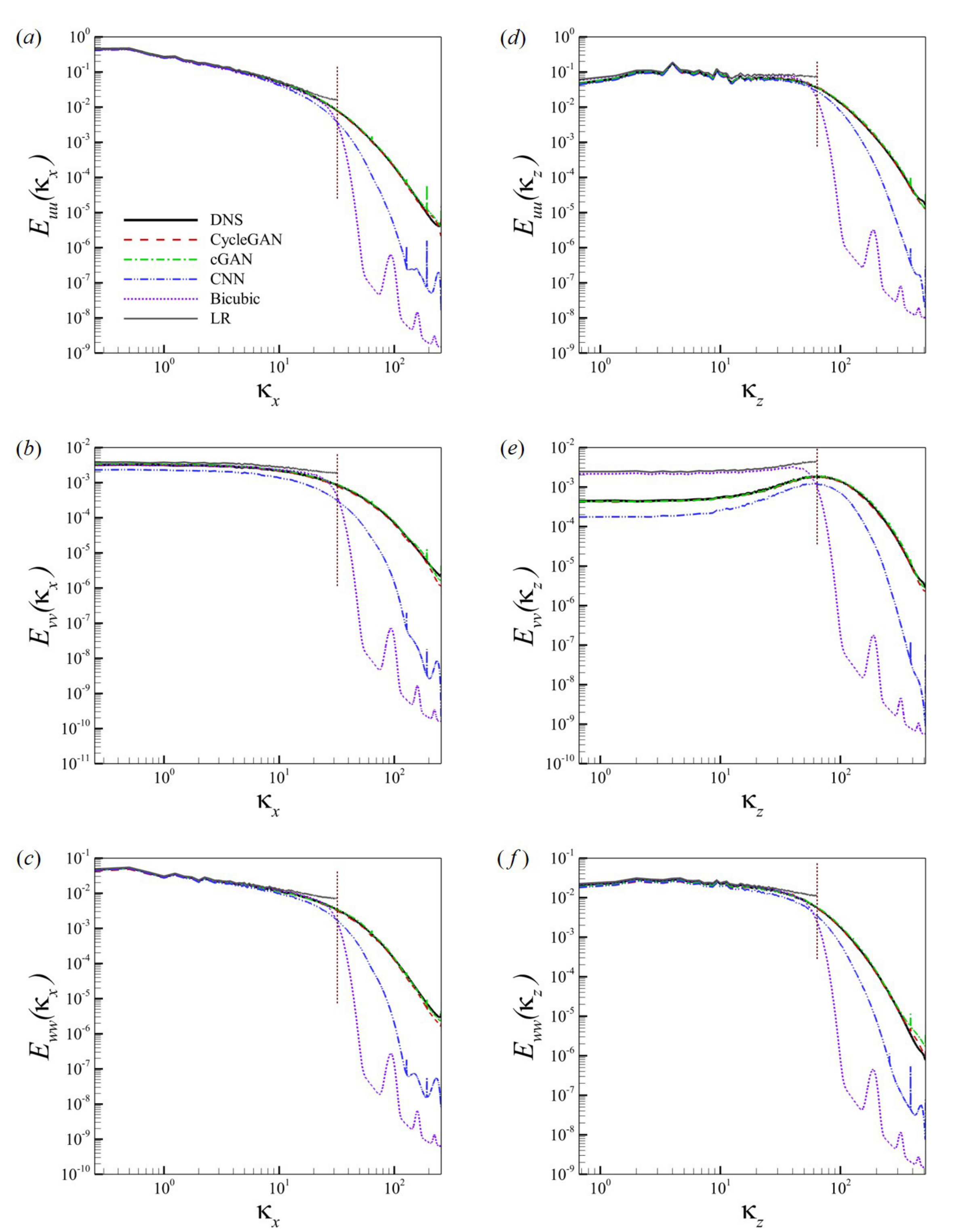}}
	\caption{1D energy spectra of reconstructed flow field using deep-learning models at $y+=15$. Streamwise energy spectra for (\textit{a}) streamwise velocity, (\textit{b}) wall-normal velocity and (\textit{c}) spanwise velocity; spanwise energy spectra for (\textit{d}) streamwise velocity, (\textit{e}) wall-normal velocity, and (\textit{f}) spanwise velocity.}
	\label{channel.energy}
\end{figure}
The streamwise and spanwise energy spectrum of each velocity component are shown in Figure \ref{channel.energy}(\textit{a,b,c}) and (\textit{d,e,f}), respectively. Statistics are averaged using test datasets. In the streamwise energy spectrum, bicubic interpolation and CNN could not reproduce DNS statistics at high wave numbers. On the other hand, cGAN accurately expressed DNS statistics at high wavenumbers. Despite using unpaired data, CycleGAN, an unsupervised learning model, showed similar results as cGAN. The spanwise energy spectrum showed similar results as the streamwise one. However, both low-resolution data and bicubic interpolation had higher energies than did DNS statistics at low wave numbers, as shown in Figure \ref{channel.energy}(\textit{e}). These results are closely related to the structure size of the reconstructed velocity field. In Figure \ref{channel.velocity}, the reconstructed field through bicubic interpolation includes structures larger than those observed in the DNS flow field in the spanwise direction. Notably, an artificial large-scale structure can be generated by interpolation if the measuring is not carried out with sufficient density. The predicted flow through the CNN requires overall smaller energy than does the DNS statistics. Particularly, this phenomenon is prominent at a relatively high wave numbers. When the input (i.e., low-resolution field) and target (i.e., high-resolution field) are not uniquely connected, CNN tends to underestimate the energy. On the other hand, cGAN and CycleGAN can accurately describe the statistics of the DNS at both low and high wave numbers. The reconstructed flow field and statistics of the energy spectrum at $y^+=100$ show similar results (see Appendix \ref{appB}).

When using partially measured data, as with experimental situations, our model can reconstruct fully measured data in wall-bounded turbulences and probably other types. By considering the pixel loss in the unsupervised learning of the homogeneous data, the point-by-point error and phase error can be reduced effectively. Compared to cGAN, which shows excellent performance among supervised learning models, CycleGAN shows similar results despite using unpaired data. CycleGAN can reconstruct the flow fields that reflect the characteristics of each velocity component, and they are statistically similar to the target DNS.
	
\subsection{Example 3: application to large-eddy simulation data }\label{subsec:LES}
In this section, we apply CycleGAN to a more practical situation in which supervised learning is impossible, because paired data are not available. We investigate whether CycleGAN can reconstruct high-resolution flow fields having DNS-quality from LES data. For unsupervised learning, we chose the same DNS data of channel turbulence as those used in Section \ref{subsec:channel}. For LES data, we numerically solved filtered Navier--Stokes equations and collected two types of data obtained using the Smagorinsky subgrid-scale model \citep{Smagorinsky1963} and the Vreman subgrid-scale model \citep{Vreman2004}. We validated that the basic statistics of LES, such as mean and RMS profile of velocities, showed similar tendencies as those of the DNS. The detailed LES setup is given in Appendix \ref{appC}. The LES and DNS data used in the training process were 2D velocity fields ($u,v,w$) in an $x-z$ plane at $y^+=15$ at $Re_\tau=1,000$. DNS training data contained 100 velocity fields of $2,048\times 1,536$ size, and LES training data contained 10,000 velocity fields of $128\times 128$ size. LES data were collected per $\Delta t = 0.004$ non-dimensionalized by $u_\tau$ and $\delta$. The domain size of DNS was $L_x \times L_y \times L_z = 8\pi \delta \times 2\delta \times 3 \pi \delta$, and that of LES was $L_x \times L_y \times L_z = 2\pi \delta \times 2\delta \times \pi \delta$. Based on the same length scale, the resolution ratio between LES and DNS was four in both $x$ and $z$ directions. 
	
During the training of CycleGAN, input (LES) and the output size of the generator $G$ were fixed as $32\times32$ and $128\times128$, respectively. After training, the input size was not fixed, and the output had $4\times4$ higher resolution than the input. The trained model was tested using 100 LES fields independent from the training data. In Section \ref{subsec:channel}, it was confirmed that the phase shift of structures might occur in the reconstructed flow field when statistically homogeneous data are used in learning CycleGAN. This can be prevented by introducing pixel loss. Similarly, in this section, a new loss ($\mathcal{L}_{LR}$) is added to the existing loss function (Equation $\ref{full loss}$) in unsupervised learning. The added loss term is defined as 
\begin{equation}
	\mathcal{L}_{LR} =  \lambda\mathbb{E}_{x\sim P_{X}}[\frac{1}{N_{p}}\sum_{i=1}^{N_p}(x(p_i)-\mathcal{I}G(p_i))^2],
	\label{eq:LR}
\end{equation}
where $x$ is the LES data used as input data, $\mathcal{I}$ is top-hat filter operation, $\mathcal{I}G(p_i)$ is the filtered flow field after reconstruction through $G$ (the same size as the input data), and $p_i$ and $N_p$ are the  position and size of the low-resolution field, respectively. The value of $\lambda$ is 10. This loss is proposed based on the assumption that the filtered flow field has a similar distribution as LES data. Using this, we expect that the small-scales will be reconstructed while the phase of the large-scale structures is maintained.
	
\begin{figure}
	\centerline{\includegraphics[width=1\columnwidth]{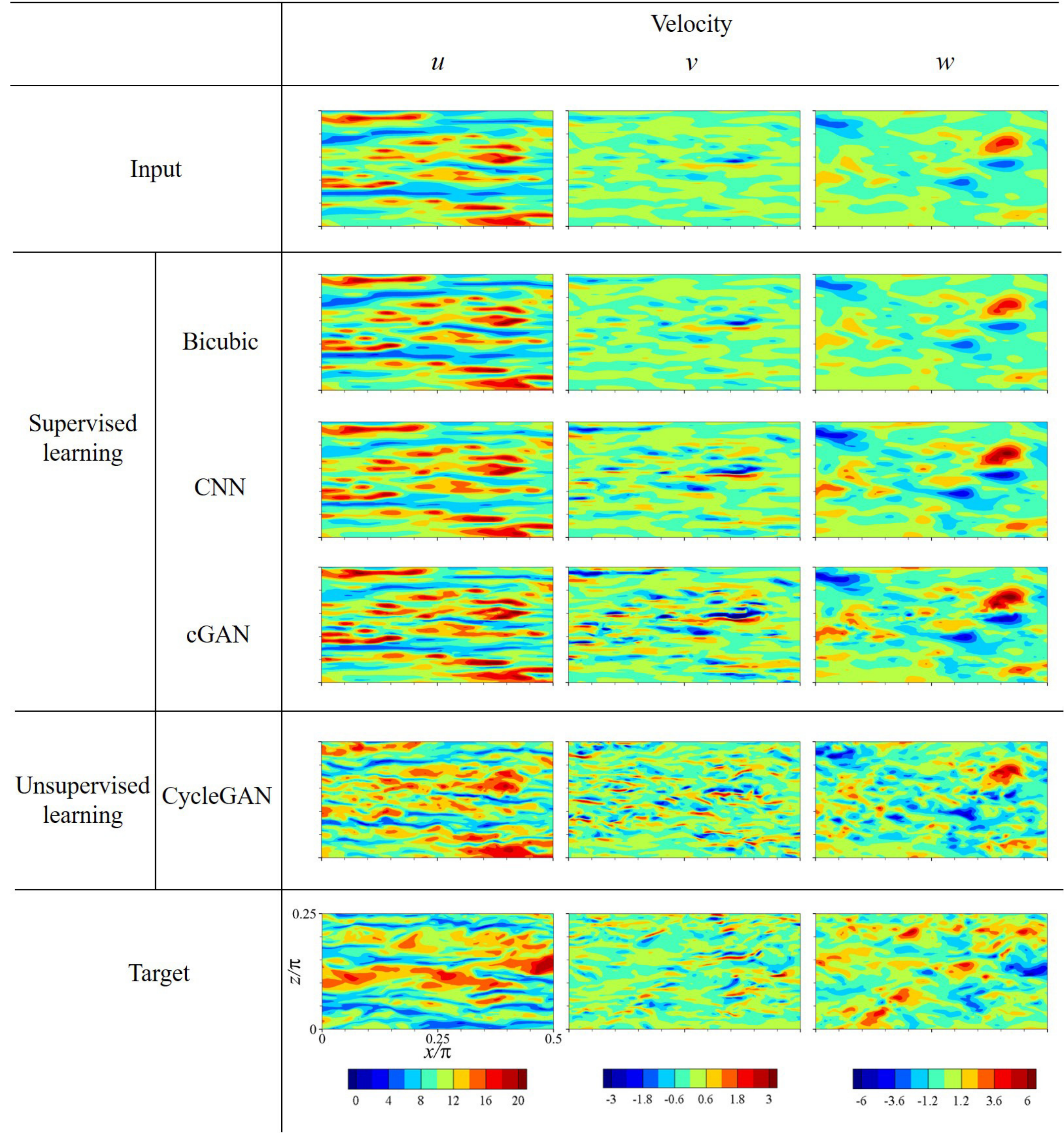}}
	\caption{Reconstructed instantaneous velocity fields $(u, v, w)$ at $y^+=15$ obtained by various deep-learning models. During the training of CycleGAN, the flow field of LES with  theVreman model is used, and a new LES field having the same model is used for testing.}
	\label{LES.velocity}
\end{figure}

For the first test, we used data obtained using the same subgrid-scale model as those used for training, the Vreman model. An example of the reconstructed velocity fields with DNS resolution from the LES data is shown in Figure \ref{LES.velocity}. Learning both LES and DNS data was possible only with the cycle-structured GAN. For comparison, we presented velocity fields reconstructed by supervised learning models (i.e., CNN and cGAN) that were trained using filtered DNS data. As shown in Figure \ref{LES.velocity}, only the CycleGAN could reconstruct a flow field that captured the features of each velocity component of the DNS field. Meanwhile, CycleGAN maintains the large-scale structure observed in LES data. On the other hand, neither bicubic interpolation nor the CNN could generate small-scale structures of DNS at all. Although cGAN demonstrated the best performance among supervised learning models (Sections \ref{subsec:isotropic} and \ref{subsec:channel}), it provided a slightly better flow field than did the CNN, and it was difficult to acknowledge that the generated fields correctly reflected DNS characteristics. In particular, the structure of the wall-normal velocity did not represent the tilted feature in the spanwise direction frequently observed in DNS data. This clearly suggests that the deep-learning model that trained the fDNS will not likely work well in the super-resolution reconstruction of LES data. We assumed that this would occur, because the DNN is overfitted to the training environment and becomes very sensitive to the input data distribution. Therefore, to successfully apply a deep-learning model to LES, an environment and a methodology capable of learning LES data are required. CycleGAN indeed meets this requirement for our super-resolution reconstruction of LES data. 

\begin{figure}
	\centerline{\includegraphics[width=1\columnwidth]{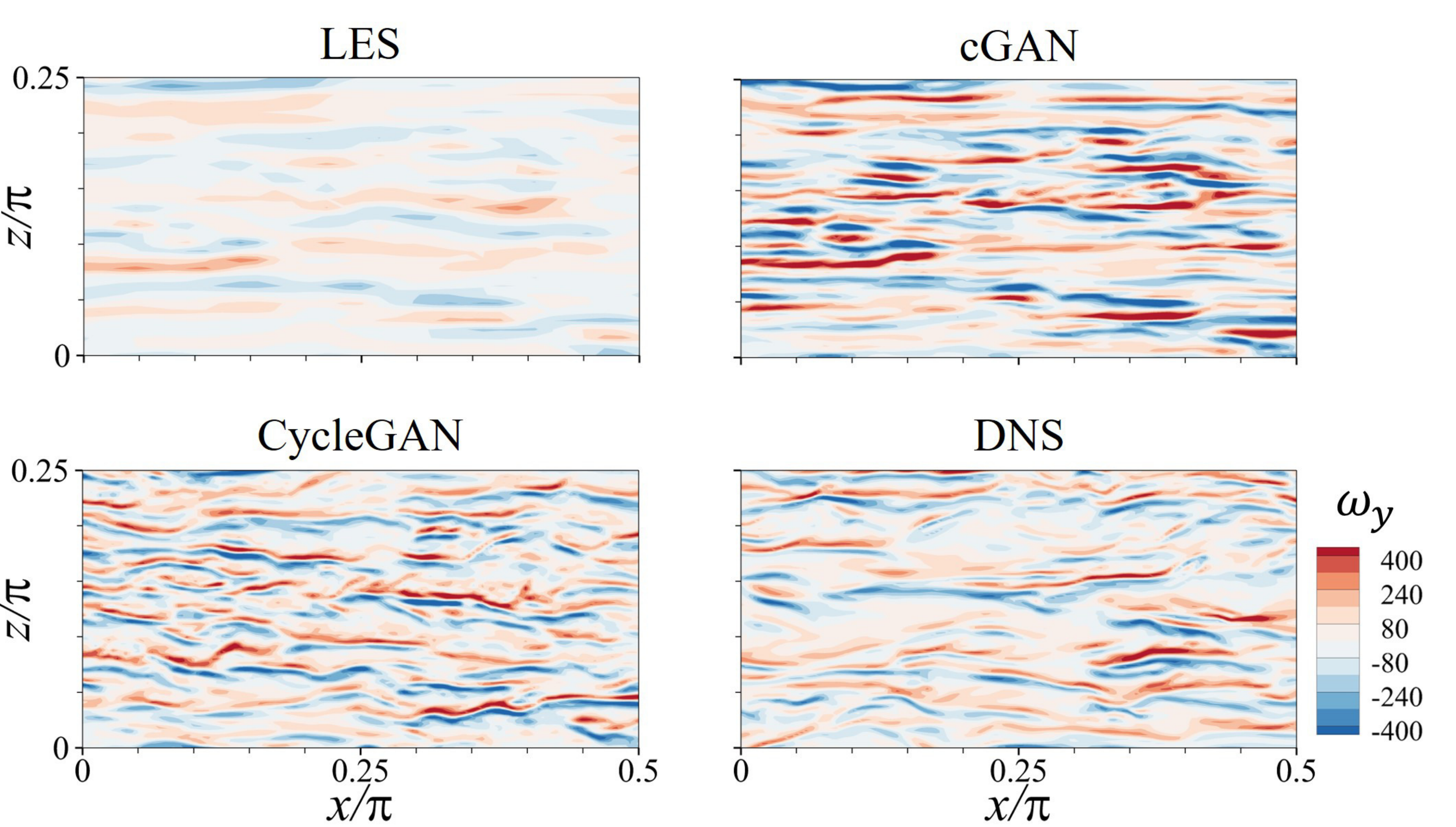}}
	\caption{Instantaneous wall-normal vorticity field calculated from reconstructed velocity fields at $y^+=15$ by cGAN and CycleGAN with input LES and target DNS fields.}
	\label{LES.vorticity}
\end{figure}
The wall-normal vorticity field obtained from the reconstructed velocity is presented with the vorticity of the input LES field in Figure \ref{LES.vorticity}. Because vorticity is not directly considered in training, it can be a good measure for assessing the performance of learning. The vorticity of LES data was much weaker than that of DNS, and the cGAN reconstructed the vorticity field much stronger than that of DNS. The thin streaky structures of vorticity found in DNS data were not captured by cGAN. Recall that cGAN was trained using filtered DNS, not LES, because the cGAN required paired data. However, structures of the vorticity field reconstructed by CycleGAN showed a striking similarity to that of the DNS. CycleGAN indeed showed an ability to accurately reconstruct the high-order component obtained through differentiation.

\begin{figure}
	\centerline{\includegraphics[width=1\columnwidth]{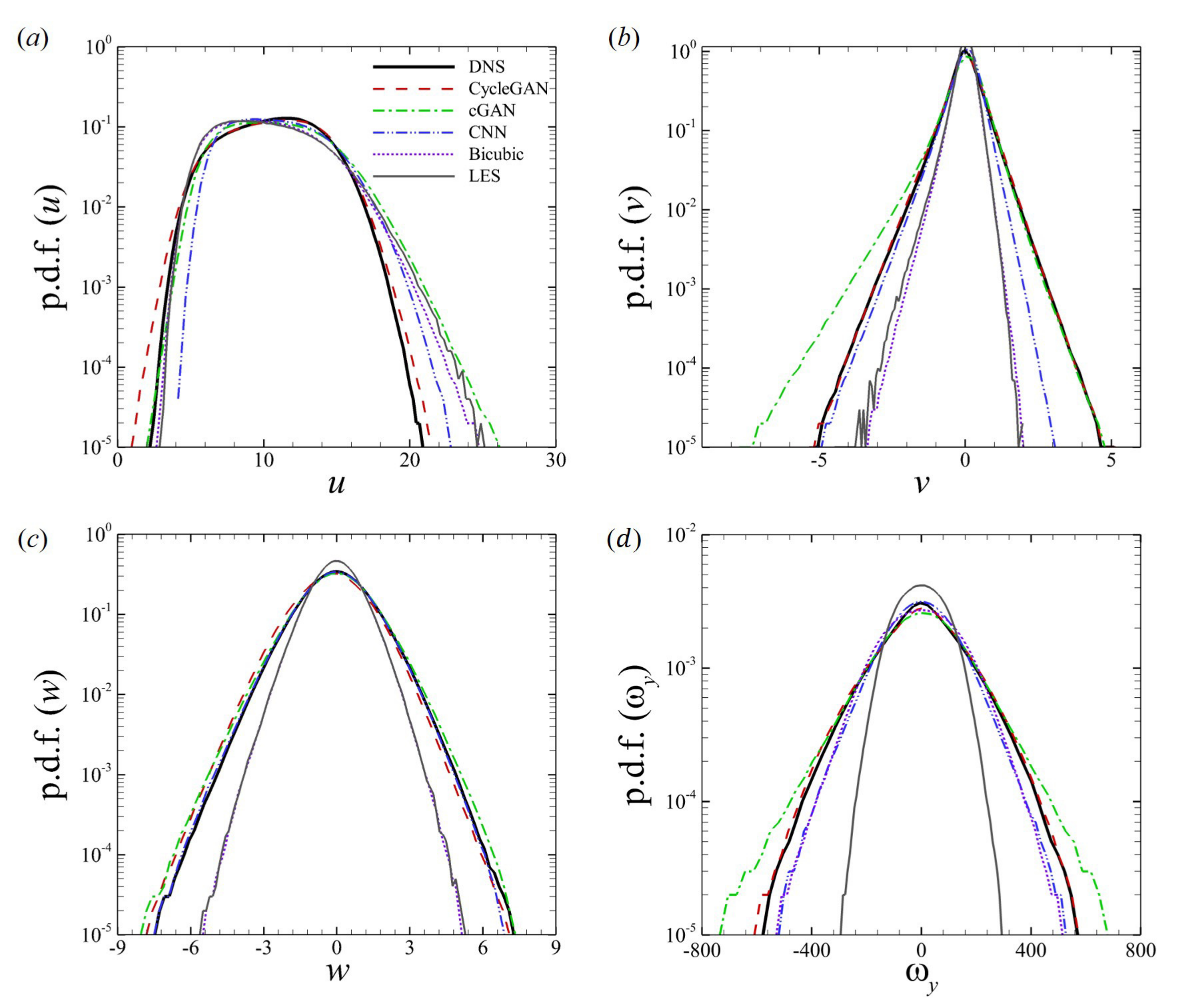}}
	\caption{Probability density function of (\textit{a}) streamwise velocity, (\textit{b}) wall-normal velocity, (\textit{c}) spanwise velocity, and (\textit{d}) wall-normal vorticity obtained from reconstructed velocity fields by various deep-learning models.}
	\label{LES.PDF}
\end{figure}

$P.d.f.$ of the reconstructed velocities and wall-normal vorticity, $\omega_{y}$, is presented in Figure \ref{LES.PDF}. The velocity $p.d.f.$ obtained by bicubic interpolation was similar to the LES statistics, not the DNS statistics. Additionally, the $p.d.f.$ by either CNN or cGAN did not well approximate that of DNS, except for the spanwise velocity, especially in a high-magnitude range. cGAN overestimated the range of the vorticity, as shown in Figure \ref{LES.PDF}(\textit{d}). On the other hand, the $p.d.f.$ of all velocity components and the vorticity by CycleGAN closely reproduced that of DNS, except only for the low-speed range of streamwise velocities. Additional quantitative statistics obtained from test data, including mean, RMS, Reynolds stress, skewness, and flatness, are presented in Table \ref{LES}. Likewise, bicubic interpolation had nearly the same value as did LES in all statistics, except for vorticity statistics. The supervised learning models (i.e., CNN and cGAN) generally showed results closer to the DNS statistics than did bicubic interpolation. However, they differed significantly from DNS in skewness of velocities and RMS of wall-normal velocity and vorticity. On the other hand, CycleGAN shows similar results to DNS in all statistics.

\begin{table}
	\begin{center}
		\begin{tabular}{cccccccc}
				\multirow{2}{*}{} & \multirow{2}{*}{} & \multicolumn{6}{c}{Deep-learning models} \\ 
				\cmidrule(lr){3-8} & & LES & Bicubic & CNN & cGAN & CycleGAN & DNS \\ \hline
				\multirow{3}{*}{Mean} & $u$ & 10.255 & 10.397 & 10.878 & 10.833 & 10.749 & 10.725 \\[1mm]
				& $v$ & 0 & 0 &0.001 & -0.002 & 0.002& 0 \\[1mm]
				& $w$ & -0.015 & -0.014 & -0.021 & -0.011 & -0.0193 & -0.007 \\	\hline
				\multirow{4}{*}{RMS} & $u$ & 3.115 & 3.119 & 2.826 & 3.102 & 2.940 & 2.831 \\[1mm]
				& $v$ & 0.324 & 0.339 & 0.487 & 0.652 & 0.598 & 0.570 \\[1mm]
				& $w$ & 0.945 & 0.966 & 1.298 & 1.361 & 1.341 & 1.292 \\[1mm]
				& $\omega_y$ & 90.0 &147.5 &142.1 & 185.7 & 168.5 & 161.6  \\	\hline
				Reynolds stress & $uv$ & -0.543 & -0.531 & -0.714 & -0.990 & -0.745 & -0.661 \\ \hline
				\multirow{4}{*}{skewness} & $u$ & 0.451 & 0.444 & 0.338 & 0.341 & -0.056 & -0.050\\[1mm]
				& $v$ & -1.033 & -1.049 &-0.852& -1.078 & -0.231 & -0.212\\[1mm]
				& $w$ & -0.035 & -0.019 &0.040 & -0.340 & -0.081 & -0.005\\	[1mm]
				& $\omega_y$ & 0.001 & -0.017 & 0.001 & -0.119 & -0.081 & -0.008 \\	 \hline
				\multirow{4}{*}{flatness} & $u$ & 2.731 & 2.667 & 2.508 & 2.650 & 2.375 & 2.378\\[1mm]
				& $v$ & 7.702 & 6.871 & 6.597 & 8.809 & 6.022 & 6.620\\[1mm]
				& $w$ & 3.695 & 3.687 & 3.685 & 3.761 & 3.536 & 3.691\\[1mm]
				& $\omega_y$ & 2.738 & 3.306 & 3.826 & 4.526 & 3.570 & 3.706 \\	
		\end{tabular}
		\caption{Velocity and vorticity statistics of reconstructed flow field at $y^+=15$ obtained by various deep-learning models.}{\label{LES}}
	\end{center}
\end{table}
	
\begin{figure}
	\centerline{\includegraphics[width=1\columnwidth]{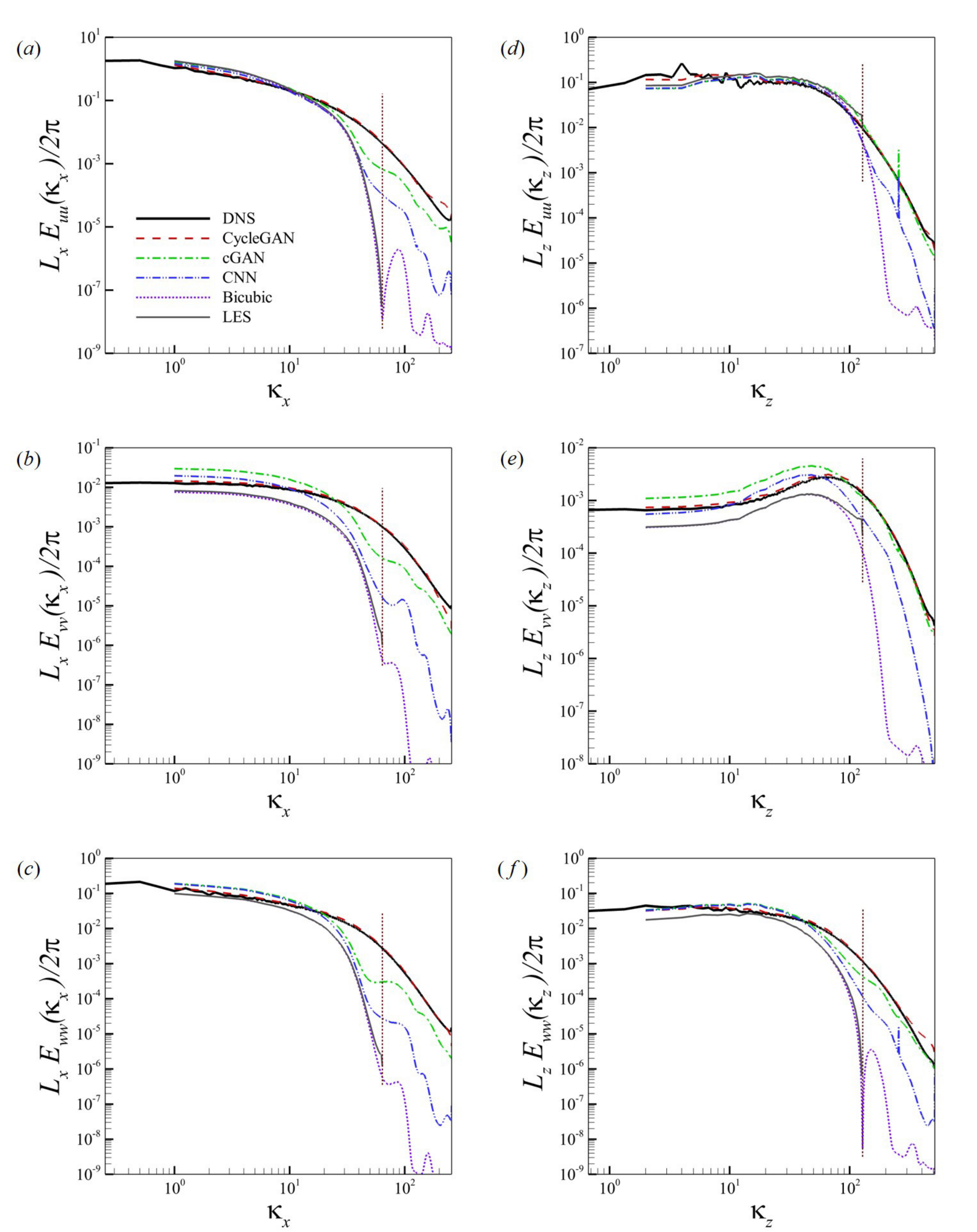}}
	\caption{1D energy spectra for reconstructed velocity field at $y+=15$. Streamwise energy spectra of (\textit{a}) streamwise velocity, (\textit{b}) wall-normal velocity, and (\textit{c}) spanwise velocity; spanwise energy spectra of (\textit{d}) streamwise velocity, (\textit{e}) wall-normal velocity, and (\textit{f}) spanwise velocity.}
	\label{LES.energy}
\end{figure}
Further assessment of learnings can be carried out with an investigation of energy spectra. The spectrum of the velocity field at $y^+=15$ is presented in Figure \ref{LES.energy}, where the streamwise and spanwise spectrum of each component of velocity are shown in Figure \ref{LES.energy}(\textit{a}),(\textit{b}), and (\textit{c}) and Figure \ref{LES.energy}(\textit{d}),(\textit{e}), and (\textit{f}), respectively. For comparison, the LES statistics used as input data are presented together, and the vertical dotted line indicates the maximum wave number of the LES. Overall, bicubic interpolation could not improve the spectrum of LES. Additionally, the supervised learning models (i.e., CNN and cGAN) tended to underestimate DNS statistics at high wave numbers. Although cGAN appeared to represent small-scale energies in the streamwise and wall-normal directions well (Figure \ref{LES.energy}(\textit{d,e})), it seemed to be a coincidence, given the flow field comparison in Figure \ref{LES.velocity}. It is noteworthy that, for the wall-normal velocity (Figure \ref{LES.energy}(\textit{b,e})), the supervised learning models could cause large errors, even at low-wave numbers. On the other hand, CycleGAN showed excellent performance in recovering overall DNS statistics via the learning of unpaired LES and DNS data. In particular, even when there was a difference in energy between LES and DNS at low wave numbers, CycleGAN reproduced DNS statistics properly (Figure \ref{LES.energy}(\textit{b,e})). This indicates that the supervised learning models were sensitive to input data. Thus, it was difficult to expect good performance for new data having distributions different from the training data. Meanwhile, CycleGAN reconstructed the flow field with the statistics of the target field by reflecting the statistical differences between LES and DNS.
	
\begin{figure}
	\centerline{\includegraphics[width=1\columnwidth]{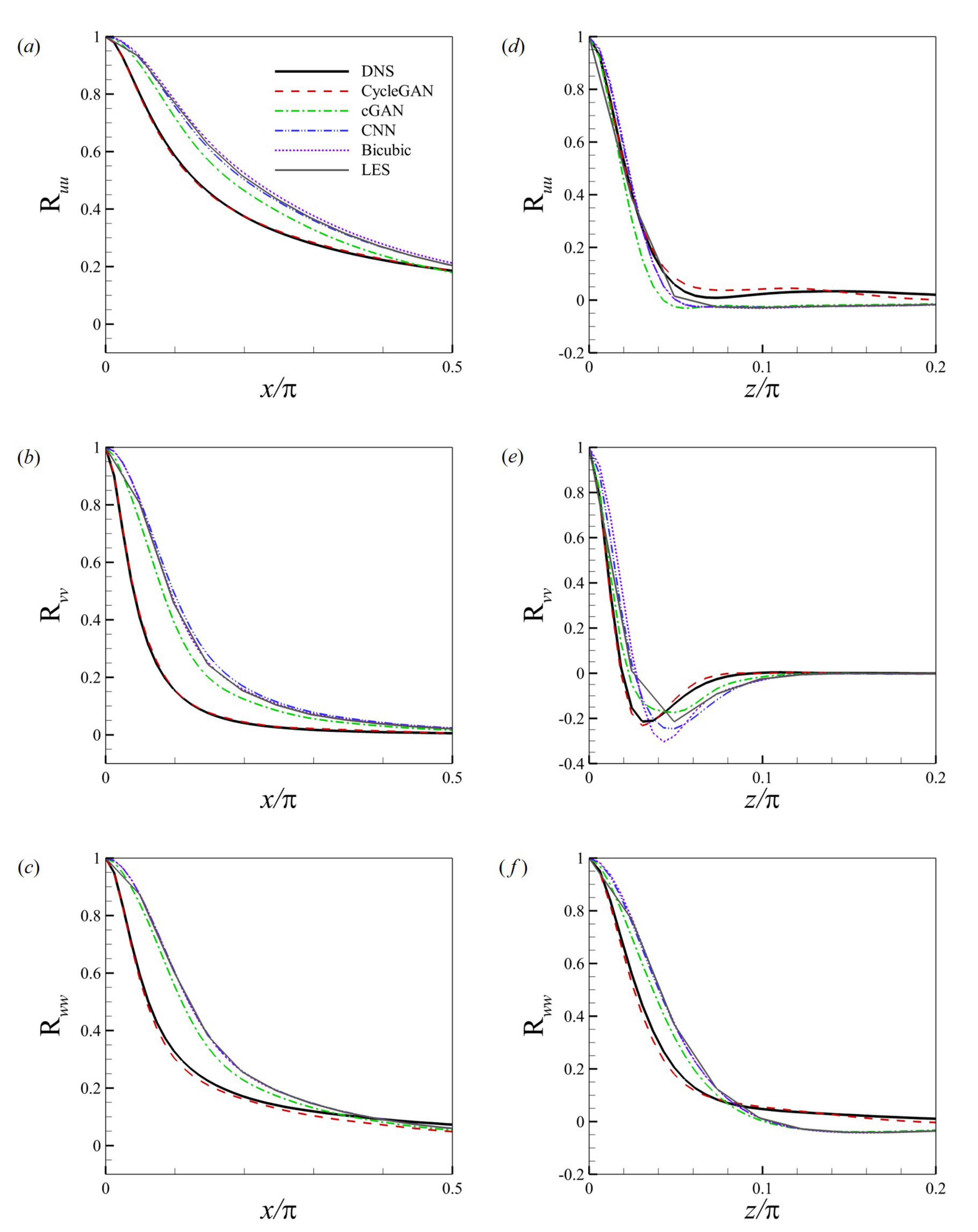}}
	\caption{Two-point correlation for reconstructed velocity field from LES data at $y^+=15$: (\textit{a}) streamwise velocity, (\textit{b}) wall-normal velocity, and (\textit{c}) spanwise velocity in streamwise statistics; (\textit{d}) streamwise velocity, (\textit{e}) wall-normal velocity, and (\textit{f}) spanwise velocity in spanwise ones.}
	\label{LES.correlation}
\end{figure}
We also checked two-point correlations of the reconstructed velocity field in Figure \ref{LES.correlation}, in which the streamwise and spanwise correlations for various learnings were compared in Figure \ref{LES.correlation}(\textit{a}),(\textit{b}), and (\textit{c}), and figure \ref{LES.correlation}(\textit{d}),(\textit{e}), and (\textit{f}). The distribution of all correlations by CycleGAN was nearly indiscernible from that of DNS. On the other hand, prediction by bicubic interpolation and supervised learning models (i.e., CNN and cGAN) could not mimic the DNS statistics, and they tended to be close to the LES statistics. The two-point correlation of LES data was mostly higher than that of DNS data, because the near-wall structures elongated in the streamwise direction were less tilted in the spanwise direction. The reconstructed flow fields using supervised learning models could not capture this tilted feature, as shown in Figure \ref{LES.velocity}. Additionally, as shown in Figure \ref{LES.correlation}(\textit{d,e,f}), the minimum position of the spanwise correlation by bicubic interpolation and supervised learning models was quite different from that of the DNS. This position as known to be related to the spacing of high- and low-speed streak and diameter of streamwise vertical structures \citep{Kim1987}. Therefore, this means that the flow field reconstructed by supervised learning models contained non-physical structures. However, the accurate statistics of CycleGAN indicated that it could represent physically reasonable structures.

\begin{figure}
	\centerline{\includegraphics[width=1\columnwidth]{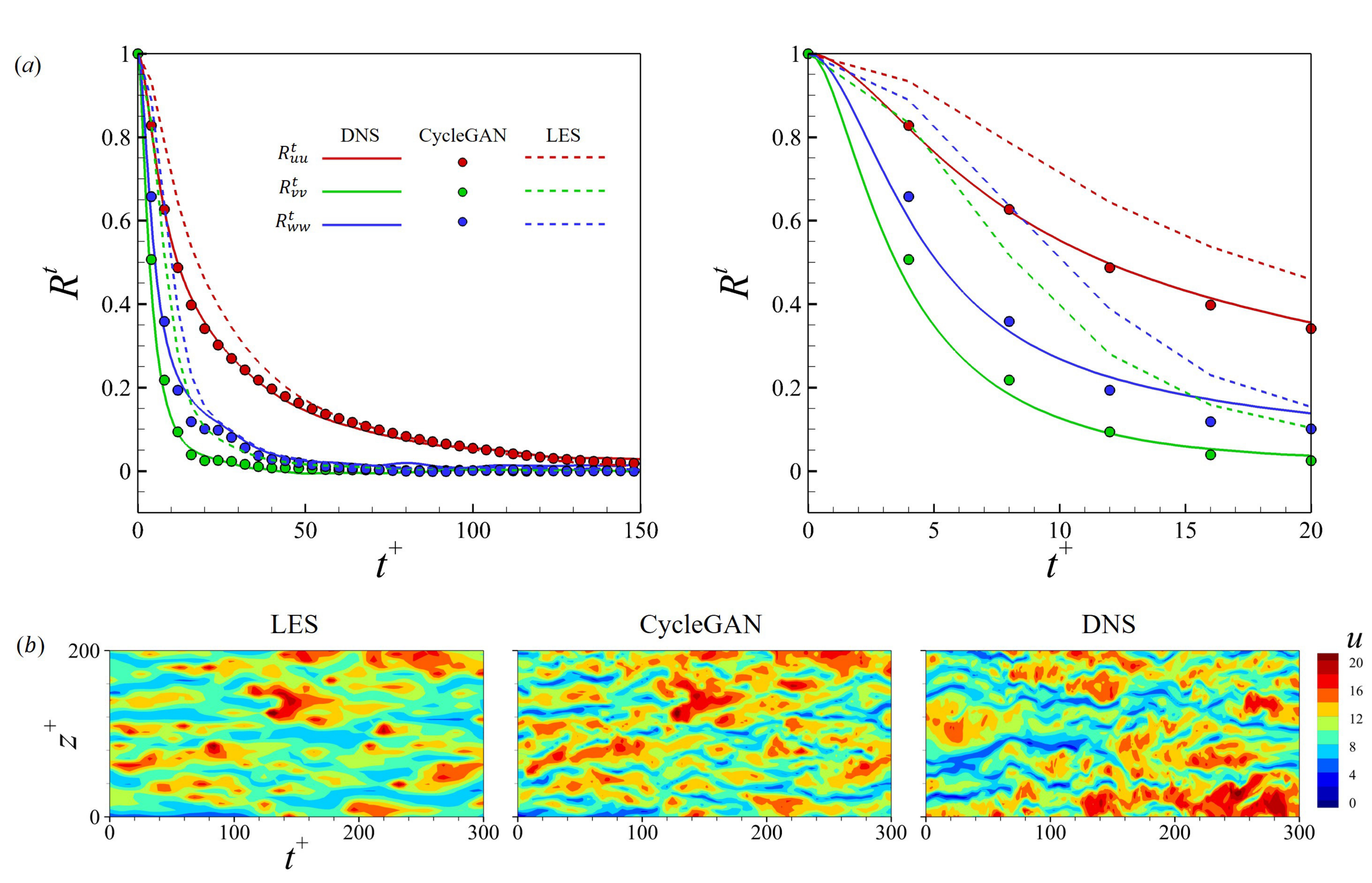}}
	\caption{(\textit{a}) Temporal correlation velocities for LES, CycleGAN, and DNS. Right panel of (\textit{a}) is a magnified view of left one near the origin. (\textit{b}) Temporal behavior of the streamwise velocity field at $y^+=15$.}
	\label{LES.time_correlation}
\end{figure}

Our CycleGAN model successfully reconstructed the super-resolution field of instantaneous the low-resolution turbulence field obtained by filtering, pointwise measurement, and independent LES. However, temporal information was not considered during training. Here, we investigate whether the trained network can reproduce correct temporal behavior of turbulent field by testing our model in the reconstruction of temporally consecutive fields. Temporal correlation, defined as $R^t_{V_iV_i}(p) = \left<V_i(t)V_i(t+p) \right>$ of the reconstructed fields by CycleGAN, is demonstrated with that of DNS and LES data in Figure \ref{LES.time_correlation}, where $\left< \right>$ denotes an average operation. Clearly, the correlation by CycleGAN recovered that of the DNS, which was quite different than that of the LES in the early period shown in the right panel of figure \ref{LES.time_correlation}(\textit{a}). In Figure \ref{LES.time_correlation}(\textit{b}), the spatio-temporal behavior of the streamwise velocity field shows that the structures by CycleGAN were tilted in the spanwise direction, resembling that of the DNS. This is an encouraging result, because it showed that the temporal information was not necessary for successful training of super-resolution reconstruction.

\begin{figure}
	\centerline{\includegraphics[width=1\columnwidth]{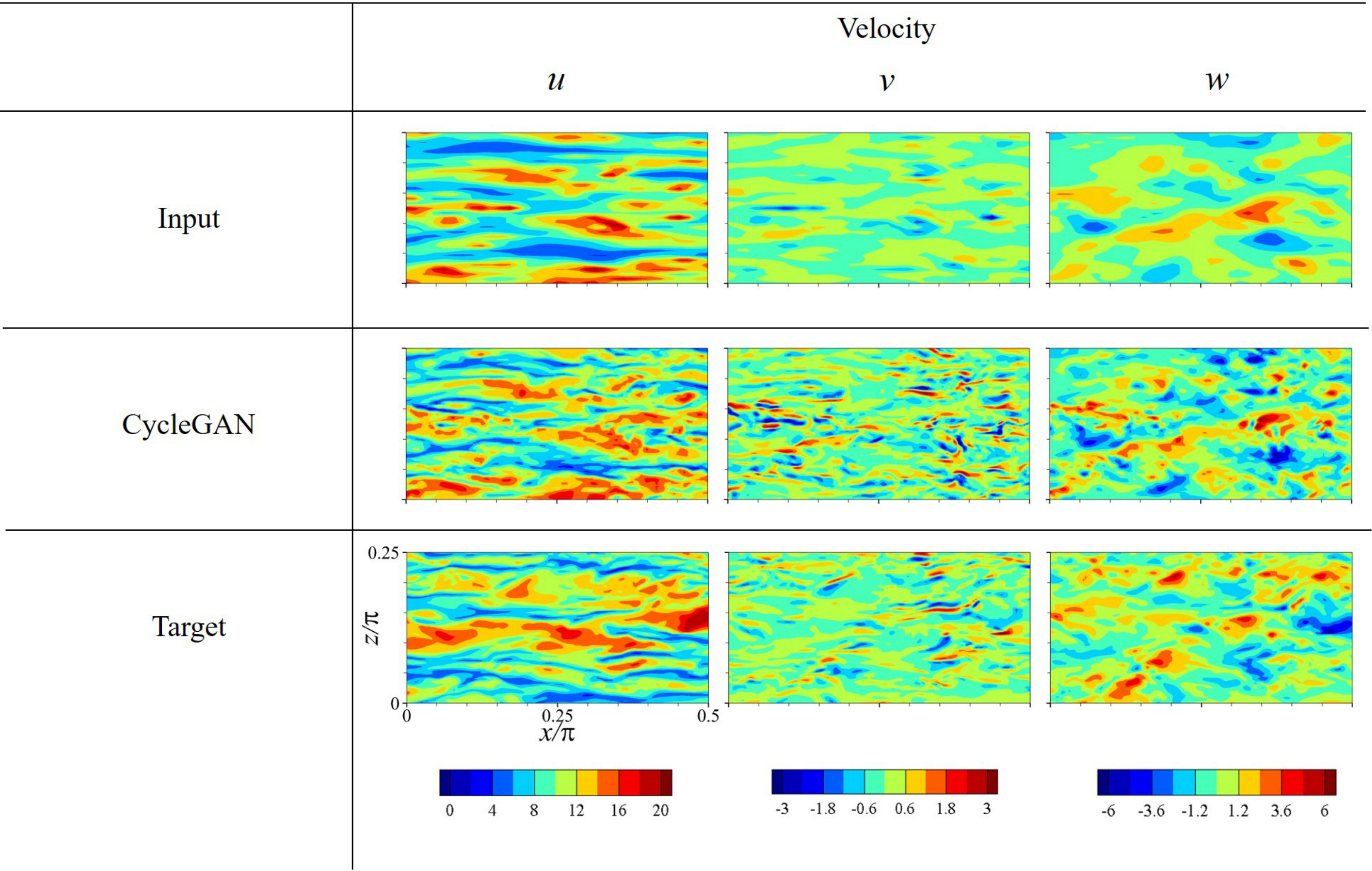}}
	\caption{Reconstructed instantaneous velocity field $(u, v, w)$ at $y^+=15$ from testing the CycleGAN model against data obtained using a different LES model. The LES model used in the training process is the Vreman model, and the test data contain the velocity field from the Smagorinsky model.}
	\label{LES1.velocity}
\end{figure}

\begin{figure}
	\centerline{\includegraphics[width=1\columnwidth]{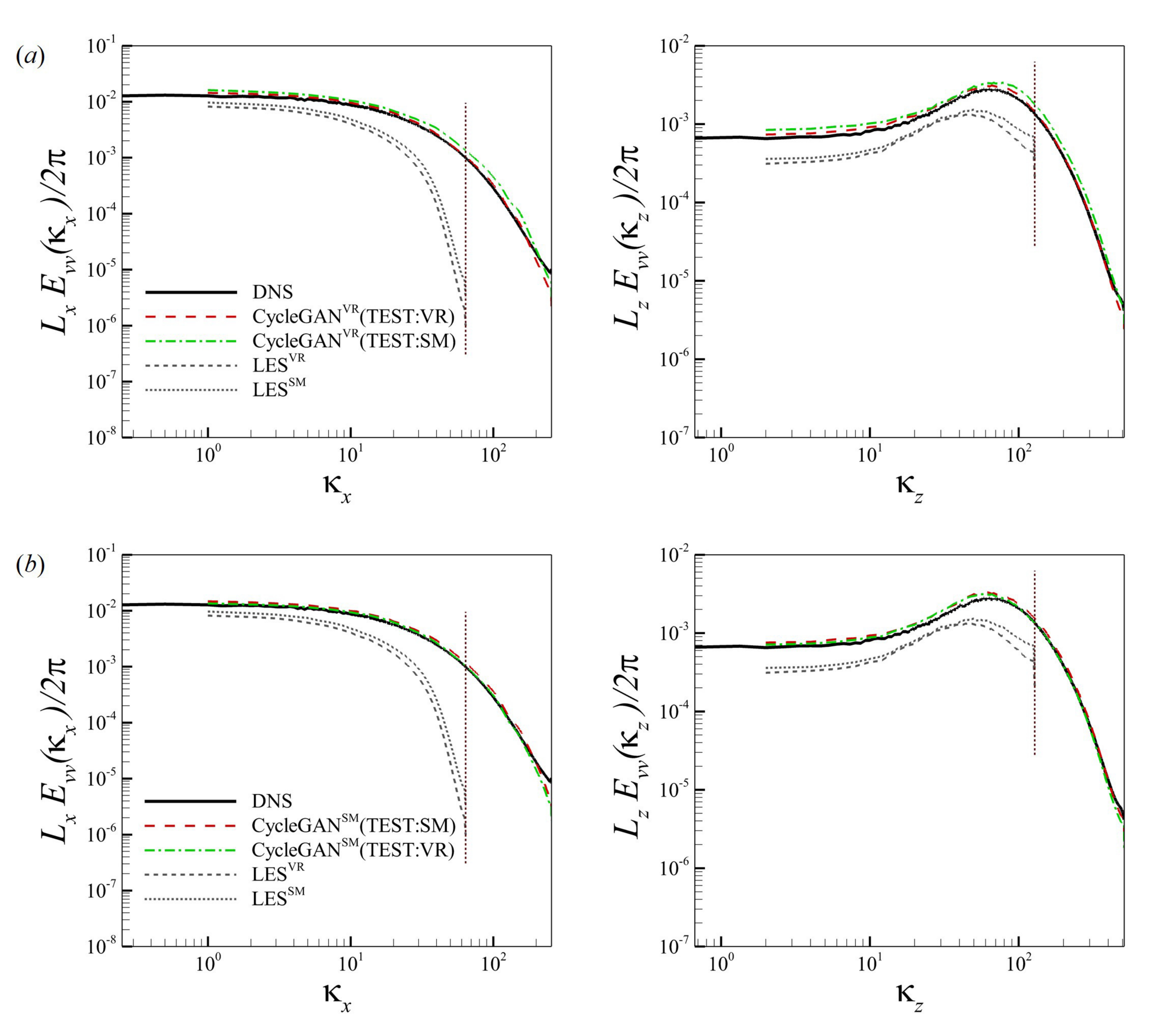}}
	\caption{1D energy spectra for reconstructed wall-normal velocity field at $y^+=15$. (a) Test results of the CycleGAN trained using LES data obtained from the Vreman subgrid-scale model; (b) test results of the CycleGAN trained using LES data obtained from the Smagorinsky subgrid-scale model. CycleGAN$^{\textnormal{VR}}$ and CycleGAN$^{\textnormal{SM}}$ denote CycleGAN models trained using LES data of Vreman and Smagorinsky models, respectively. LES$^{\textnormal{VR}}$ and LES$^{\textnormal{SM}}$ are LES input data from the Vreman and Smagorinsky models, respectively.}
	\label{LES.energy2}
\end{figure}

Finally, we investigated the performance of CycleGAN in a test against different kinds of input LES data. Our CycleGAN was trained and tested using the input LES data obtained by the Vreman subgrid-scale model. Here, we tested this CycleGAN for the input LES data obtained by a different subgrid-scale model: the Smagorinskly model. As shown in Figure \ref{LES1.velocity}, the CycleGAN reconstructed the velocity fields that reflect the characteristics of DNS, despite the use of data from different LES models. Quantitatively, the comparison of the 1D energy spectra of the reconstructed wall-normal velocity between LES input data obtained by the Vreman model and the Smagorinsky model clearly demonstrates that both yielded nearly the same distribution as that of DNS, although that from the Smagroinsky LES data showed a slight overestimation for most wave numbers, as shown in Figure \ref{LES.energy2}(\textit{a}). As a cross validation, CycleGAN was trained using LES data obtained by the Smagorinsky model, and it was tested with LES data obtained by the Vreman model. As shown in Figure \ref{LES.energy2}(\textit{b}), CycleGAN reproduced DNS-quality reconstructed fields for both input data. Recall that the cGAN, which was trained using filtered DNS data, could not well-reconstruct DNS-quality data from LES data. This clearly shows the advantage of unsupervised learning in a situation where paired data are not available.

\section{Conclusion}\label{Conclusion}

We presented an unsupervised learning model that adopted CycleGAN to reconstruct small-scale turbulence structures when low- and high-resolution fields were unpaired. To investigate the performance of CycleGAN, an interpolation method (i.e., bicubic interpolation) and supervised learning models (i.e., CNN and cGAN) were considered. The supervised learning models were trained using paired low- and high-resolution data. We considered homogeneous isotropic turbulence and a turbulent channel flow where paired data existed. Finally, we demonstrated super-resolution reconstruction with DNS characteristics from LES fields in a channel flow where only unpaired data exist.

First, we investigated the performance of various learning models for different resolution ratios, $r$, between input fields obtained by applying a top-hat filter to DNS data, and we output DNS fields in homogeneous isotropic turbulence. Bicubic interpolation and CNN did not well reconstruct the small-scale structures. The energy spectrum and vorticity $p.d.f.$ statistics yielded by bicubic interpolation and CNN were rather similar to those of low-resolution input data. On the other hand, cGAN showed excellent ability to recover small scales, even for large $r$. Similarly, our CycleGAN provided excellent performance in the reproduction of energy spectrum and $p.d.f.$ of vorticity, despite using unpaired data.

Next, we assessed the performance of the super-resolution reconstruction of anisotropic turbulence in a limited measurement environment. Low-resolution data were extracted by pointwise measurement of high-resolution DNS data at $y^+=15$ and $100$ of channel turbulence. The phase shift of structures in the reconstructed flow field from unsupervised learning in spatially homogeneous data was eliminated by introducing pixel loss, which is the point-by-point MSE of the measured information. As predicted, bicubic interpolation and CNN did not reconstruct small-scale structures, similar to the previous example. On the other hand, the cGAN showed high accuracy in reconstruction, reflecting the characteristics of the DNS. The flow fields reconstructed through CycleGAN were good as those provided by the cGAN, and their statistics were similar to those of DNS.

Finally, CycleGAN was applied to a more practical problem of reconstructing a flow field with DNS quality from LES data unpaired from DNS data. Supervised learning models (i.e., CNN and cGAN) were trained using filtered DNS data, because paired data were not available. Trained CNN and cGAN did not produce small scales, and the reconstructed flow field had different structures from the DNS data. All statistics, including the $p.d.f.$ of velocity and vorticity, the energy spectrum, and the two-point correlations, showed a completely different distribution from those of DNS. On the other hand, CycleGAN effectively reconstructed the flow field that reflected the structures of each velocity and vorticity observed in DNS. All statistical quantities produced by CycleGAN were consistent with those of DNS. The temporal behavior of turbulent fields were correctly captured by the reproduced fields obtained by the application of CycleGAN to consecutive LES fields. Finally, we applied CycleGAN to LES data using a different subgrid-scale model that was not used for training, and it showed excellent performance. 

There are several remaining issues that should be considered in future works. First, low-resolution data lack information required to uniquely reproduce high-resolution data in general. When the resolution ratio is large, different high-resolution data can be generated, depending on the initial value of trainable parameters in the network, and trained networks randomly map only one of many possible high-resolution solutions. However, this might be unavoidable because of the intrinsic nature of turbulence and its strong sensitivity to small disturbances. Second, when low-resolution data are provided on an irregular mesh rather than on a uniform mesh, it is inappropriate to apply the current convolution operation. A technique, such as graph CNN \citep{Kipf2016}, could be used to resolve this problem. Third, the present study assumed that there was a sufficient amount of high-resolution data for training. There might be some situations in which only limited amount of high-resolution data or even no data are available. Good solutions should include data augmentation using symmetry, physics-informed NNs that impose constraints of governing equation (continuity or momentum equations) \citep{Raissi2019}, and physical constraints added to the NN \citep{Mohan2020}. Fourth, the current study was limited to the super-resolution reconstruction of the instantaneous 2D flow field, but a consideration of the temporal behavior or 3D information of the flow might yield better or more efficient reconstructions. For example, it is possible to account for temporally successive data by adding a discriminator that considers temporal effects \citep{Xie2018,Kim2020}. Finally, the analysis of trained network was difficult because of the large number of parameters. \citet{Kim2020b}, for example, identified that the gradient map of trained model could be used to extract the physics implied in the training data. This progress, with respect to super-resolution reconstruction, might help identify a nonlinear relationship between large-scale structures and small-scale ones. 

We have shown that super-resolution reconstruction of turbulence using CycleGAN is possible in situations where paired data are not available. We expect that the proposed network will be of great assistance to LES modeling, including the production of pair data for the development of subgrid-scale models and synchronizations for model evaluation. Furthermore, our model can be utilized to support high-resolution reconstruction of measurement data, such as PIV \citep{Rabault2017,Cai2020}, synchronization of different experiments, removal of experimental noise, and data assimilation \citep{Leoni2020}. 

\vspace{0.3in}
\noindent
{\bf {Acknowledgments}}

This work was supported by a National Research Foundation of Korea (NRF) grant funded by the Korea government (MSIP) (2017R1E1A1A03070282).

\appendix
\section{Network architecture and hyperparameters of deep learning model}\label{appA}
\begin{figure}
	\centerline{\includegraphics[width=1\columnwidth]{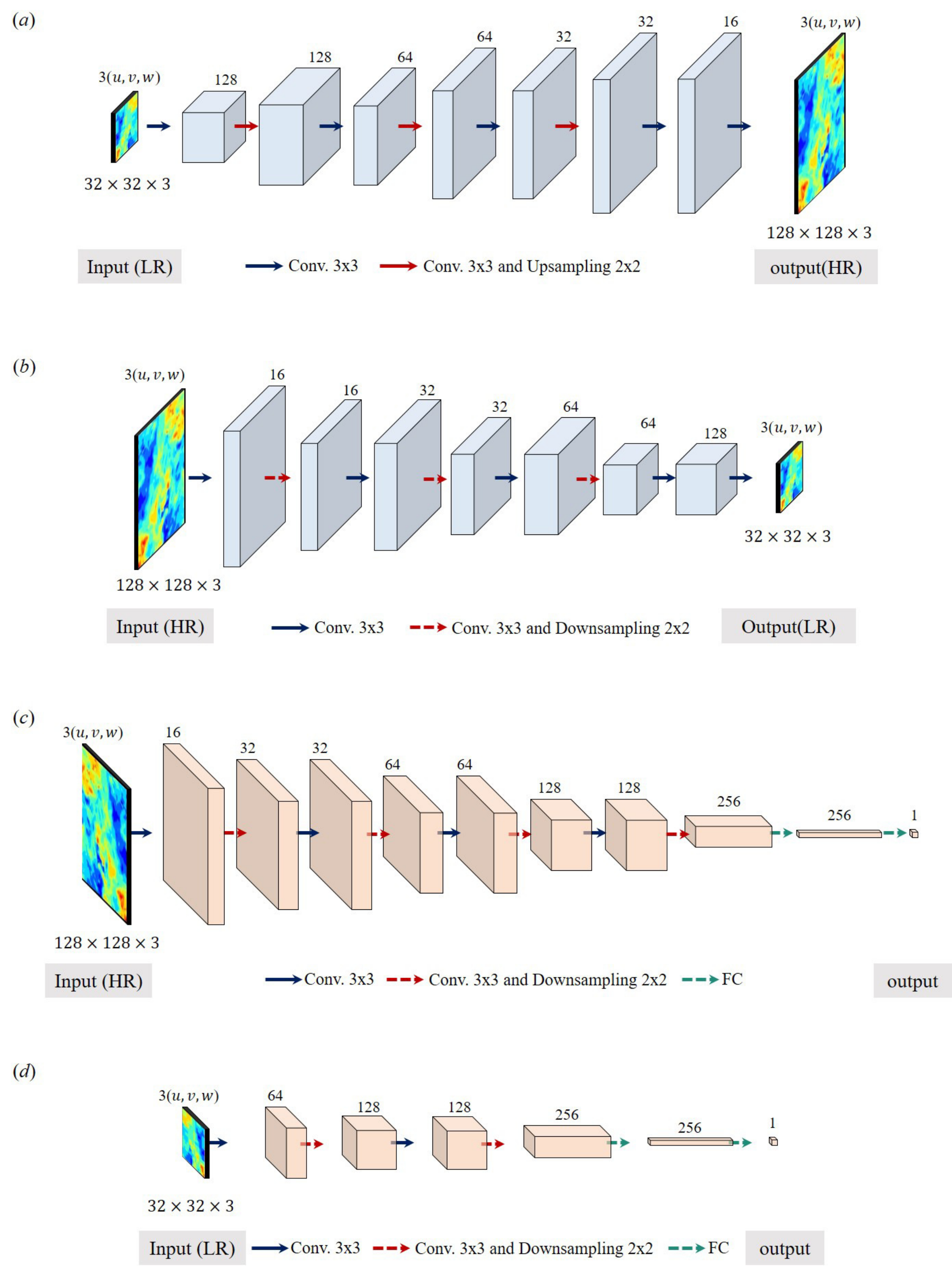}}
	\caption{Network architecture of generators and discriminators of CycleGAN for resolution ratio $r=8$. (\textit{a}) generator G. (\textit{b}) generator F. (\textit{b}) discriminator $D_Y$. (\textit{d}) discriminator $D_X$.}
	\label{architecture}
\end{figure}

Figure \ref{architecture} shows the network architecture of components of CycleGAN when the resolution ratio was eight. A CycleGAN consisted of two generators (Figure \ref{architecture}(\textit{a}) $G$ and (\textit{b}) $F$) and two discriminators (Figure \ref{architecture}(\textit{c}) $D_X$ and (\textit{d}) $D_Y$). The objective of learning was to obtain $G$ that could reconstruct the high-resolution turbulent field. $G$ comprised convolution (Conv. in Figure \ref{architecture}) and up-sampling operations, repetitively. $F$, $D_X$, and $D_Y$ comprised convolution and down-sampling operations. In $D_X$ and $D_Y$, a fully connected layer (FC in Figure \ref{architecture}) was additionally used to yield one value. The size of the discrete convolution operation was fixed at $3\times3$. During this process, a padding was used to maintain the size of input data. Zero padding was used during the training process, and periodic padding was used during testing to automatically satisfy the periodic boundary condition. Nearest-neighborhood interpolation and average pooling were used for up- and down-sampling with $2\times2$ size, respectively. Following the convolutions and fully connected layers, except for the last layer, a nonlinear activation function (Leaky ReLU in Equation \ref{LeakyReLU}) was applied. 
Depending on the resolution ratio, the number of convolution layers, up- and down-sampling operations in the network changed slightly. Trainable parameters were randomly initialized \citep{He2015}. During training, learning rate, batch size, and total iterations were 0.0001, 16, and 500,000, respectively. The Adam optimizer \citep{kingma2014adam} was used for minimizing and maximizing the objective function. There was room for improvement via changes in architecture, such as batch normalization \citep{ioffe2015batch}, residual networks \citep{he2016deep}, and fine-tuning of hyperparameters. 

The supervised learning models (i.e., CNN and cGAN), which were used for comparison, comprised the same generator network as $G$ of CycleGAN.
The discriminator of cGAN was nearly the same as $D_Y$ of CycleGAN, except for the channel size of the input. The same hyperparameters were used for CNN, cGAN, and CycleGAN, except for the learning rate and total iterations of the CNN. The initial learning rate of CNN was 0.0005, and we reduced it by 1/5 when the validation error did not decrease. 

\section{Test in the outer-region of wall-bounded turbulent flows}\label{appB}
\begin{figure}
	\centerline{\includegraphics[width=1\columnwidth]{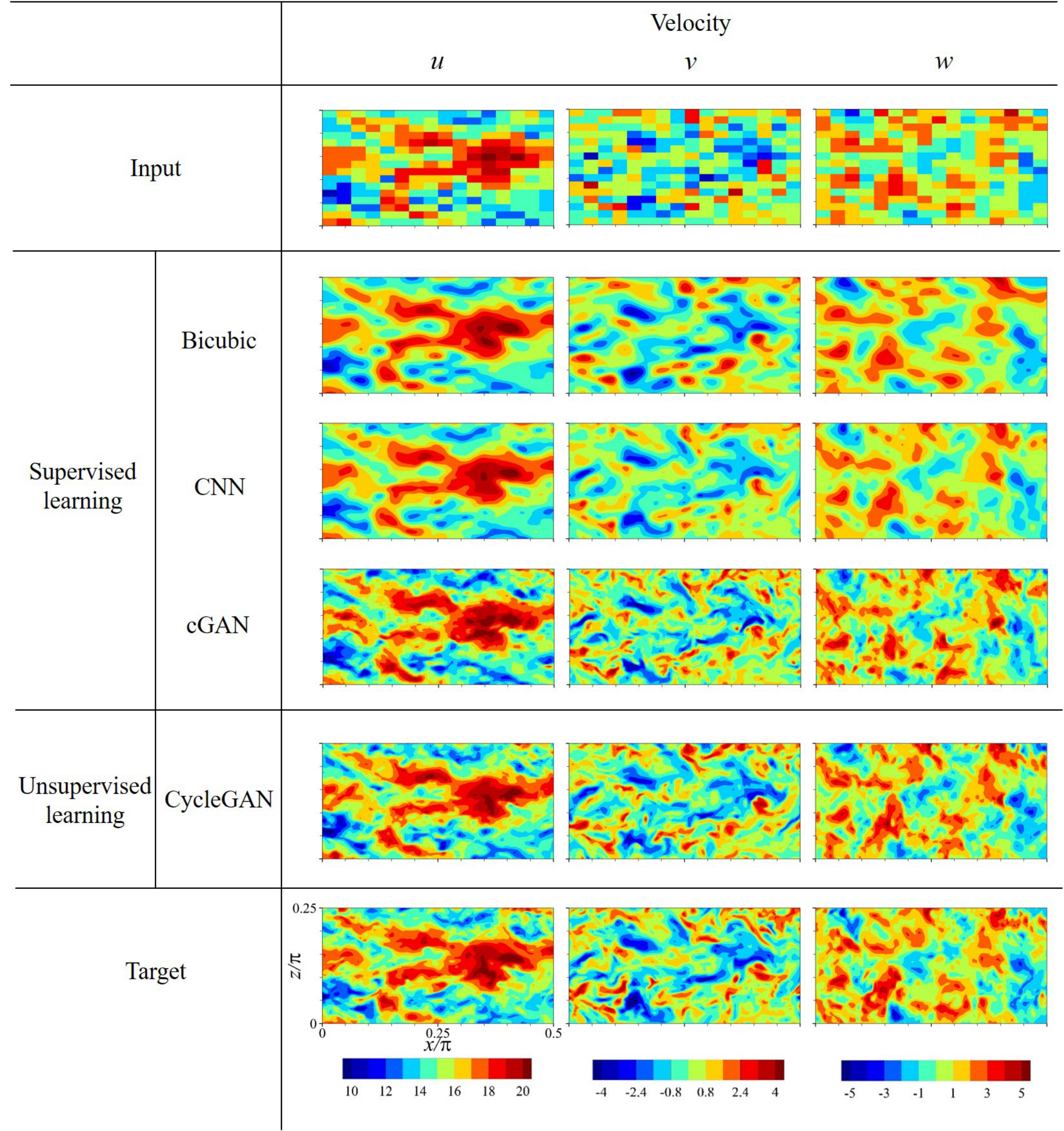}}
	\caption{Reconstructed instantaneous velocity fields at $y^+=100$ obtained by various deep learning models.}
	\label{channel.y100velocity}
\end{figure}
\begin{figure}
	\centerline{\includegraphics[width=1\columnwidth]{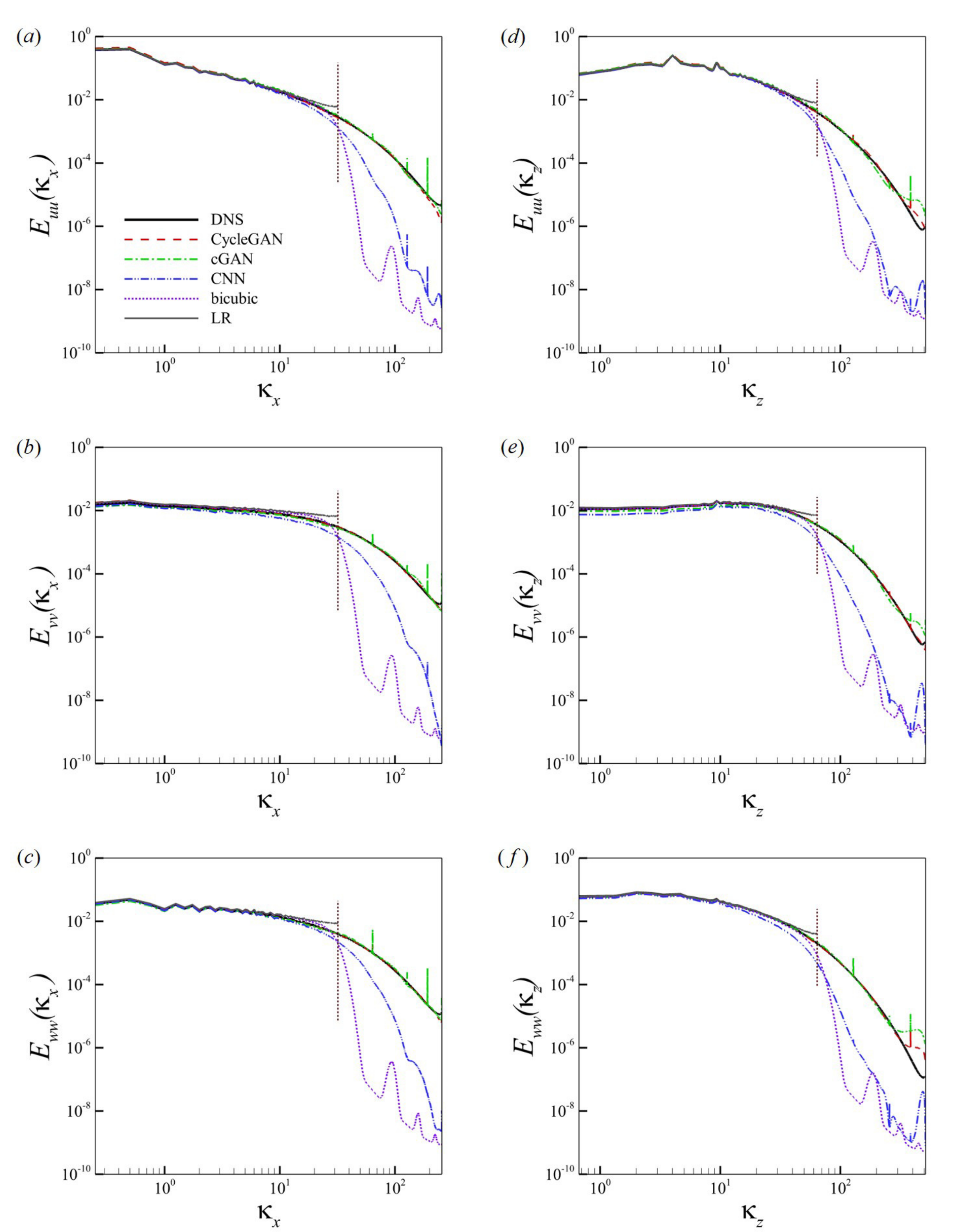}}
	\caption{One-dimensional energy spectra for reconstructed velocity field at $y^+=100$: Streamwise energy spectra of (\textit{a}) streamwise velocity, (\textit{b}) wall-normal velocity and (\textit{c}) spanwise velocity. Spanwise energy spectra of (\textit{d}) streamwise velocity, (\textit{e}) wall-normal velocity and (\textit{f}) spanwise velocity.}
	\label{channel.y100energy}
\end{figure}

In section \ref{subsec:channel}, we applied CycleGAN to the reconstruction of the velocity fields ($u, v, w$) from partially measured data at $y^+=15$ and $100$. Considering that the input was pointwise measurement data, we additionally used point-by-point pixel loss during training. The phase of the high wave-number components in the reconstructed velocity field at $y^+=15$ was more accurate than that at $y^+=100$, as shown in figure \ref{channel.phase}. The reason might be that fluctuation intensity in the near-wall region was stronger than that of the outer-region. However, at $y^+=100$, the reconstructed velocity field of CycleGAN was as accurate as that of cGAN, which showed the best performance among supervised learning models, as shown in Figure \ref{channel.y100velocity}. The bicubic interpolation and CNN captured only large-scale structures, compared with DNS. In 1D energy spectra of reconstructed velocity fields (Figure \ref{channel.y100energy}), our model showed excellent performance, similar to DNS and cGAN. There was only a slight error with the DNS for a few specific wave numbers. The error was related to the up-sampling scheme in generator $G$. The error \textcolor{red}{could be} avoided by changing the nearest-neighborhood interpolation using only linear data \citep{Karras2018}. On the other hand, the statistics of the bicubic interpolation and CNN did not follow those of the DNS at high wave numbers. These results indicate that the CycleGAN was good enough to replace supervised learning models, which require paired datasets.
		
\section{Validation of large eddy simulation}\label{appC}
\begin{figure}
	\centerline{\includegraphics[width=1\columnwidth]{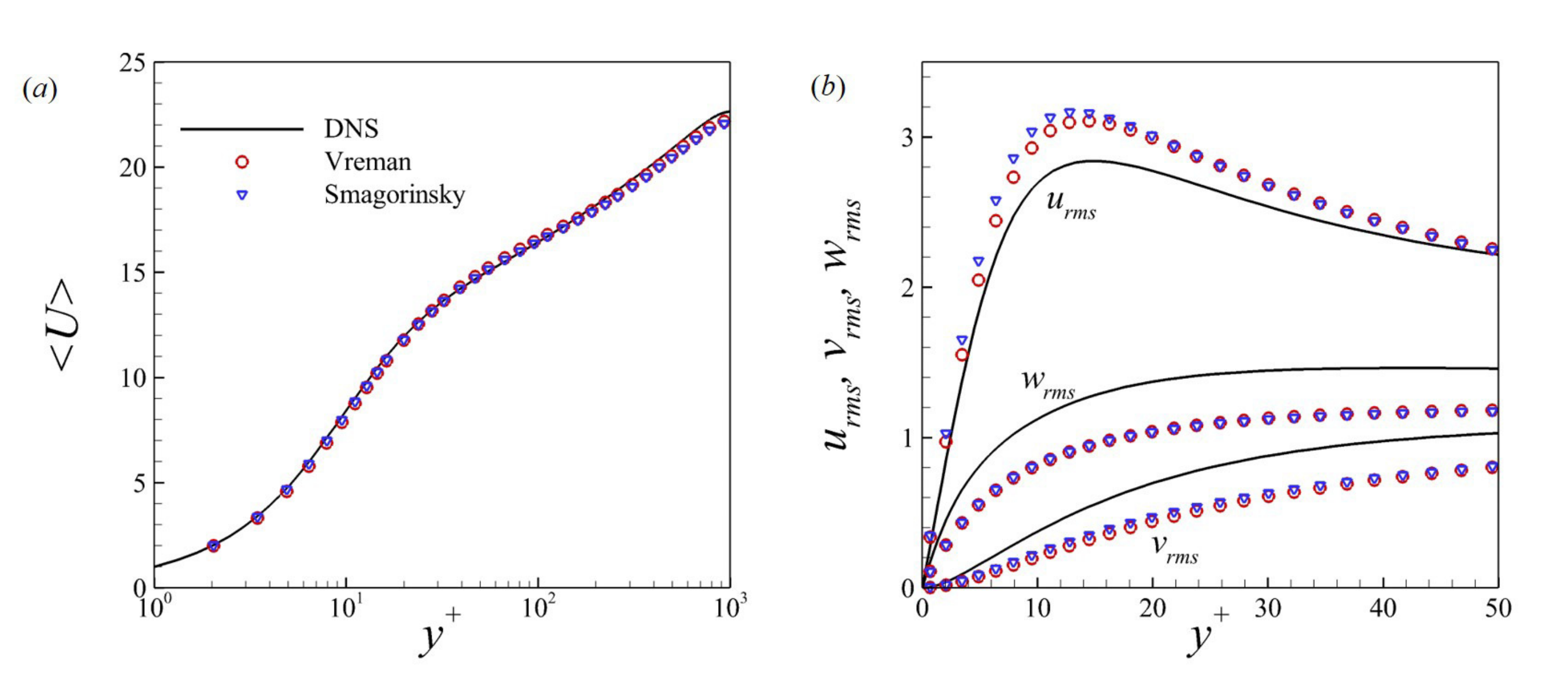}}
	\caption{(\textit{a}) Mean velocity profile in wall units, and (\textit{b}) RMS velocity profiles obtained by LES with the Vreman and Smagorinsky models.}
	\label{LES.validation}
\end{figure}

For the development of an unsupervised learning model, we required LES data, which was obtained by carrying out a large-eddy simulation of turbulent channel flow. A periodic boundary condition was imposed in the streamwise and spanwise directions. The constant mean pressure gradient drove a mean flow in the streamwise direction. The boundary layer was developed using a no-slip boundary condition at the top and bottom walls. Governing equations were those of filtered incompressible Navier--Stokes equations, which can be written as follows:
\begin{eqnarray}
{\partial \bar{u}_i \over \partial x_i}  & = & 0, \\
{\partial {\bar{u}_i} \over \partial t} +  {\partial \bar{u}_j\bar{u}_i  \over \partial x_j} & = & -{\partial \bar{p} \over \partial x_i} + {1 \over Re_\tau}{\partial^2{\bar{u}_i} \over \partial x_j \partial x_j}-{\partial \tau_{ij}  \over \partial x_j}.
\end{eqnarray} 
Equations were made dimensionless using the friction velocity, $u_\tau$, and the channel half-width, $\delta$. Here, $\bar{u}_i$ was the filtered velocity, and $\tau_{ij}$ was the subgrid-scale stress that should be modeled. We used two kinds of subgrid-scale models: Smagorinsky \citep{Smagorinsky1963} and Vreman \citep{Vreman2004}. Furthermore, the Van Driest damping, which multiplies the subgrid-scale stress by $(1-e^{-y^+/25})^2 $, was applied to the Smagorinsky model. We controlled the Smagorinsky constant, $C_{s}$, to fit the mean profile of the LES to that of the DNS. As a result, $C_{s}=0.17$ for both models. The third-order hybrid Runge--Kutta scheme was used for time integration \citep{Rai1991}, and the second-order central difference scheme was used for spatial discretization. We distributed a uniform grid in the horizontal direction, and a non-uniform grid with a hyperbolic tangent function in the wall-normal direction. We carried out LES with both subgrid-scale models using the same grid resolution of $128\times 256\times 128$ and the same domain size of $2\pi \delta \times 2\delta \times \pi \delta$ at $Re_{\tau}=1,000$. The resolution ratio, $r$, between our LES and the DNS of the JHTDB at the same Reynolds number as four for in both steamwise and spanwise directions. The time-interval, $\Delta t = 0.0004$, which was non-dimensionalized with $u_\tau$ and $\delta$. The time-averaged mean and RMS profiles are given in Figure \ref{LES.validation}. Although there is a slight gap in the RMS profile, owing to the low grid-resolution, the trend of statistics was consistent with that of the DNS. For training, we collected 10,000 velocity fields in the $x-z$ plane at $y^+=15$. The time-interval between temporally successive fields was $\Delta t = 0.004$. For testing, we used new data sufficiently far from the training data.

\bibliographystyle{jfm}
\bibliography{ref}
		
\end{document}